\documentclass[a4,11pt]{article}
\usepackage{ifpdf}
\usepackage[margin=1in]{geometry}
\usepackage[utf8]{inputenc}
\usepackage[T1]{fontenc}
\usepackage[english]{babel}
\usepackage{times}

\usepackage{authblk}
\usepackage{booktabs}
\usepackage{mathtools}
\usepackage{amsthm} 
\usepackage{amssymb}
\usepackage{enumitem} 
\usepackage{tikz}
\usepackage{array}
\usepackage{footnote}
\usepackage{xcolor}
\usepackage{graphicx}
\usepackage{adjustbox}
\usepackage[figuresleft]{rotating}

\usepackage{tabularx}
\usepackage{longtable}
\usepackage{multirow}
\usepackage{multicol}

\usepackage{placeins}
\usepackage{float}

\usepackage{makecell}
\usepackage{pifont} 
\usepackage{xspace} 

\usepackage{algorithm}
\usepackage{algpseudocode}

\usepackage{microtype} 
\usepackage[none]{hyphenat} 
\definecolor{lime}{HTML}{A6CE39}
\DeclareRobustCommand{\orcidicon}{
	\begin{tikzpicture}
	\draw[lime, fill=lime] (0,0) 
	circle [radius=0.16] 
	node[white] {{\fontfamily{qag}\selectfont \tiny ID}};
	\draw[white, fill=white] (-0.0625,0.095) 
	circle [radius=0.007];
	\end{tikzpicture}
	\hspace{-2mm}
}

\foreach \x in {A, ..., Z}{
	\expandafter\xdef\csname orcid\x\endcsname{\noexpand\href{https://orcid.org/\csname orcidauthor\x\endcsname}{\noexpand\orcidicon}}
}

\usepackage{pdflscape}
\usepackage{caption}
\DeclareCaptionFont{captionbetween}{\fontsize{7.4pt}{6.8pt}\selectfont}
\usepackage{setspace}

\usepackage{hyperref}
\hypersetup{pdfstartview=FitH, colorlinks=true, urlcolor=blue, citecolor=blue, linkcolor=blue,
pdftitle={Secure Aggregation for Privacy-Preserving Federated Learning on Clinical EEG Data},
pdfauthor={Pouya Rajabi and Mohsen Toorani},
pdfsubject={Privacy-preserving federated learning and secure aggregation for clinical EEG data},
pdfkeywords={secure aggregation, federated learning, clinical EEG, privacy, verifiability, privacy-preserving record linkage}
}

\title{Secure Aggregation for Privacy-Preserving Federated Learning on Clinical EEG Data}

\author{Pouya Rajabi\orcidA{} \hspace{4cm} Mohsen Toorani\orcidB{}\\
\vspace{10pt}
Department of Science and Industry Systems \\ University of South-Eastern Norway \\ Kongsberg, Norway}
\date{}

\begin{document}
\maketitle
\let\thefootnote\relax
\footnotetext{\scriptsize
Copyright \textcopyright~Authors 2026.
This preprint is distributed through arXiv under the arXiv.org perpetual, non-exclusive license. No additional reuse rights are granted except as provided by that license.
\par 
A version of this manuscript will appear in the Proceedings of the International Workshop on Hot Topics at the Intersection of Distributed Machine Learning and Security (HotDiSec~2026), co-located with ESORICS~2026.}

\setstretch{1.16}

\begin{abstract}
Federated learning enables multiple institutions to collaboratively train a shared model without exchanging their raw data. However, individual model updates are data-dependent and may reveal information about clients' local training data. This paper presents a privacy-preserving federated learning framework for clinical EEG data that uses masking-based secure aggregation as its core protection mechanism. The framework combines graph-based communication, threshold secret sharing, dropout recovery, local update clipping, an optional Bloom filter-based privacy-preserving record-linkage initialization module, and auxiliary-notary-based verifiability. It supports semi-honest and malicious aggregation settings and is implemented using the Flower federated learning framework. The secure aggregation variants are evaluated in a simulated cross-silo healthcare setting using TUH EEG-derived data under different client configurations. Under the stated assumptions, the secure variants hide individual updates from the aggregation server. The results show that these variants remain compatible with federated model training, although malicious-setting safeguards and lightweight consistency-checking mechanisms introduce additional computation, communication, and round-duration overhead. Among the proposed secure configurations, the base semi-honest variant incurs the lowest overhead; the malicious-server variants add protocol-consistency and authenticity safeguards, and the auxiliary-notary variants add lightweight aggregate-consistency checking.
\end{abstract}

\section{Introduction}
Federated learning (FL) has become a promising approach for collaborative machine learning across institutions without direct data sharing \cite{McMahanMRHA17}. Instead of centralizing data, FL allows each participant to train locally and share model updates with an aggregation server. This is especially relevant in healthcare, where sensitive clinical data are difficult to centralize because of privacy, regulatory, institutional, and cost constraints \cite{ThapaliyaOG0CP24}. Clinical electroencephalography (EEG) data illustrate this challenge, as recordings may contain sensitive patient-related information and vary across patients, devices, recording conditions, and institutions. Although FL keeps raw data local, it does not provide complete privacy protection because model updates may leak information through reconstruction and inference attacks \cite{ZhuLH19}. FL systems must also handle non-IID data, communication cost, runtime overhead, and participant dropout \cite{DingAHAL23}. These challenges motivate privacy-preserving mechanisms that can be integrated into iterative FL without exposing individual updates.

Secure aggregation (SecAgg) addresses this problem by allowing the server to recover only the aggregate of client updates, rather than the plaintext update of any individual participant \cite{BonawitzIKMMPRS17,SoNYL0AGA22}. Masking-based secure aggregation protocols are particularly attractive for FL because they hide individual updates using pairwise masks, self-masks, and secret sharing, while still allowing the server to reconstruct the final aggregate after dropout handling \cite{BonawitzIKMMPRS17,BellBGL020}. However, practical use in a clinical EEG-oriented setting requires more than the basic privacy goal. The aggregation mechanism must remain compatible with cross-silo FL, support bounded communication, tolerate client dropout, and provide additional safeguards when the server may behave maliciously or present inconsistent protocol views.

In this paper, we propose a practical secure aggregation framework for privacy-preserving federated learning on clinical EEG data. In the targeted cross-silo healthcare setting, institutional clients keep EEG data, preprocessing, and local training inside their own environments while participating in collaborative training. The proposed framework includes four secure aggregation variants for different threat and verification settings: semi-honest, malicious, semi-honest with auxiliary-notary verifiability, and malicious with auxiliary-notary verifiability. The framework uses masking-based secure aggregation and combines graph-based neighbor communication, threshold secret sharing, local update clipping, dropout recovery, and an optional Bloom filter-based privacy-preserving record-linkage initialization module \cite{SchnellBR09,VatsalanSCR17}. It uses public-key commitments, signed messages, acknowledgment evidence, and consistency checks in the malicious setting, while auxiliary-notary verifiability is used in the verifiable variants. The framework is implemented in Flower \cite{abs-2007-14390} and evaluated using EEG data derived from the TUH EEG Corpus \cite{tuh_eeg,obeid2016temple} under 10-client, 40-client, and 70-client configurations.  
The main contributions of this paper are as follows:
\begin{itemize}
    \item A Flower-based integration of masking-based secure aggregation into a cross-silo federated learning pipeline for clinical EEG data, combining sparse graph-based communication, local update clipping, threshold secret sharing, and dropout recovery.

    \item An implementation and comparison of multiple aggregation settings, including baseline federated learning, semi-honest secure aggregation, malicious secure aggregation, and auxiliary-notary-based verifiability variants.

    \item An experimental evaluation on TUH EEG-derived data under 10-client, 40-client, and 70-client configurations, examining model performance, protocol overhead, observed scaling behavior up to 70 clients, and privacy, security, and performance trade-offs.
\end{itemize}

The framework also contains an implemented, optional Bloom filter-based record-linkage initialization module. Because the TUH EEG Corpus contains no usable linkage attributes, this module is described and analyzed but is not included in the reported experimental evaluation.

\section{System Model and Threat Model}

\begin{figure}[!t]
  \centering
  \includegraphics[width=0.9\linewidth]{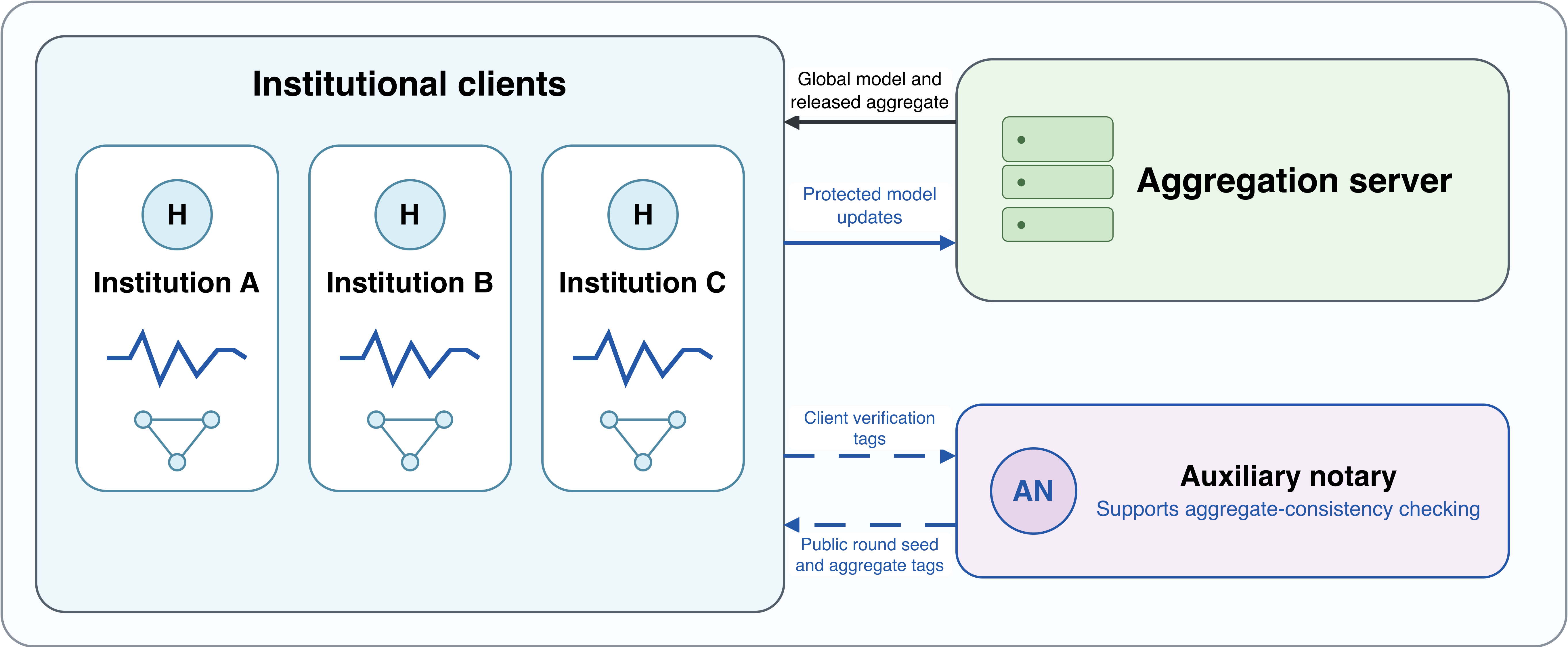}
  \caption{Cross-silo federated learning system model.}
  \label{fig:system-model}
\end{figure}

\paragraph{Cross-silo federated learning setting.}
The proposed framework aims at a cross-silo federated learning setting for clinical EEG data. A set of institutional clients collaboratively train a shared model under the coordination of a central aggregation server. Each client represents a healthcare institution that stores EEG data locally and performs preprocessing and local training inside its own environment. Clients share the same learning task and feature structure, but their local datasets may differ in size and distribution because EEG recordings vary across patients, devices, recording conditions, and institutions. 
Figure~\ref{fig:system-model} summarizes the entities and communication paths in the cross-silo setting. Raw EEG data remain at the institutional clients, which receive the global model from the aggregation server and return protected model updates. The server also releases the reconstructed aggregate to the clients for verification. To support aggregate-consistency checking, the auxiliary notary publishes a public round seed, receives verification tags from the clients, and broadcasts the corresponding aggregate tags. The clients use these values to check whether the aggregate released by the server is consistent with the tags submitted by the participating clients. The detailed verification mechanism and the protocol variants that use it are introduced in Section~\ref{sec:proposed-framework}.

\paragraph{Aggregation model.}
Let $S_r$ be the set of clients selected in round $r$, with $|S_r|=n$. The server sends the current global model $w^{(r)}$ to the selected clients, and each client $i\in S_r$ trains locally to obtain an update $x_i^{(r)}$. In baseline FedAvg, the server directly aggregates the received updates to update the global model \cite{McMahanMRHA17}. In the secure setting, $x_i^{(r)}$ is the input to the secure aggregation protocol: each client clips and masks this update before transmission, and the server should recover only the aggregate of the clients that successfully complete the round. The selected set $S_r$ is therefore refined during the protocol into active subsets such as $A_1$, $A'_2$, and later recovery sets, which determine which clients are included in the final aggregate and which masking material must be reconstructed after dropout.

\paragraph{Adversarial settings and assumptions.}
The framework considers semi-honest and malicious aggregation settings. In the semi-honest setting, the server follows the protocol, but may try to infer information from masked updates, metadata, and protocol outputs. In the malicious setting, the server may present inconsistent participation views, omit clients, request inconsistent recovery material, or return an incorrect aggregate. The framework assumes that enough clients remain active for threshold reconstruction and that raw EEG data never leave the client institutions. It does not claim protection against all attacks on federated learning, including malicious-client behavior, poisoning, backdoors, Byzantine clients, or inference from the final global model.

\section{Proposed Secure Aggregation Framework}
\label{sec:proposed-framework}
The proposed framework builds on masking-based secure aggregation to let the server compute the sum of client updates without learning individual updates. Pairwise masks are generated between neighboring clients and cancel in the aggregate, while self-masks and threshold secret sharing support dropout recovery. If a client drops out, shares held by its neighbors allow the server to remove the required masks without revealing any surviving client update. The protocol uses key agreement to derive pairwise secrets, a pseudorandom generator to expand them into mask vectors, Shamir's secret sharing to split mask seeds and private masking keys, and authenticated encryption to protect share distribution through the server relay. To reduce communication compared to dense all-to-all masking, each round uses graph-based neighbor communication. Each client communicates only with a fixed-size neighbor set $N_G(i)$, where the graph degree and reconstruction threshold are chosen so that enough surviving neighbors remain for dropout recovery while communication remains bounded. The malicious-setting extension adds Merkle public-key commitments, proof verification, signed inclusion messages, acknowledgments, and consistency checks \cite{BonawitzIKMMPRS17,BellBGL020}.

\begin{table}[!t]
\centering
\caption{Proposed secure aggregation protocol variants.}
\label{tab:protocol-variants}
\small
\setlength{\tabcolsep}{3pt}
\renewcommand{\arraystretch}{1.15}
\begin{tabular}{p{0.08\textwidth} p{0.15\textwidth} p{0.35\textwidth} p{0.35\textwidth}}
\hline
\textbf{Protocol} & \textbf{Setting} & \textbf{Main mechanisms} & \textbf{Additional protection} \\
\hline
$\Pi_1$ & Semi-honest &
Graph-based communication, pairwise masks, self-mask, Shamir sharing, encrypted shares, dropout recovery. &
Hides individual updates from an honest-but-curious server. \\
\hline
$\Pi_2$ & Malicious &
$\Pi_1$ + Merkle public-key commitment, proof verification, directed sparse graph, inclusion signatures, acknowledgments, consistency checks. &
Reduces server equivocation, forged participation claims, and unsafe recovery-share release. \\
\hline
$\Pi_3$ & Semi-honest + AN &
$\Pi_1$ + auxiliary-notary verification tags computed from public randomness and checked against the aggregate. &
Provides lightweight aggregate checking for the server-reported contributor set. \\
\hline
$\Pi_4$ & Malicious + AN &
$\Pi_2$ + the same auxiliary-notary verification mechanism used in $\Pi_3$. &
Combines malicious-setting safeguards with aggregate checking relative to the server-declared contributor set. \\
\hline
\end{tabular}
\end{table}

The experimental evaluation considers one baseline, one basic secure aggregation comparison protocol, and four proposed secure aggregation variants. The baseline FL setting is used as a performance reference and does not protect individual updates. The basic SecAgg comparison follows the masking-based secure aggregation protocol introduced by Bonawitz et al. \cite{BonawitzIKMMPRS17}, using pairwise masks, self-masks, secret sharing, and dropout recovery to hide individual updates from the aggregation server. The four proposed protocol variants are denoted by $\Pi_1$ to $\Pi_4$ and are summarized in Table \ref{tab:protocol-variants}. These variants are based on the graph-based SecAgg+ construction of Bell et al. \cite{BellBGL020}, which replaces dense all-to-all masking with sparse graph-based neighbor communication while preserving the core masking and dropout-recovery principles of secure aggregation. 
In this paper, we adapt that construction to a clinical EEG-oriented cross-silo FL pipeline and extend it with local update clipping, malicious-setting consistency checks, auxiliary-notary verifiability, and an optional privacy-preserving record linkage initialization module. 

\begin{figure}[!t]
  \centering
  \includegraphics[width=\linewidth]{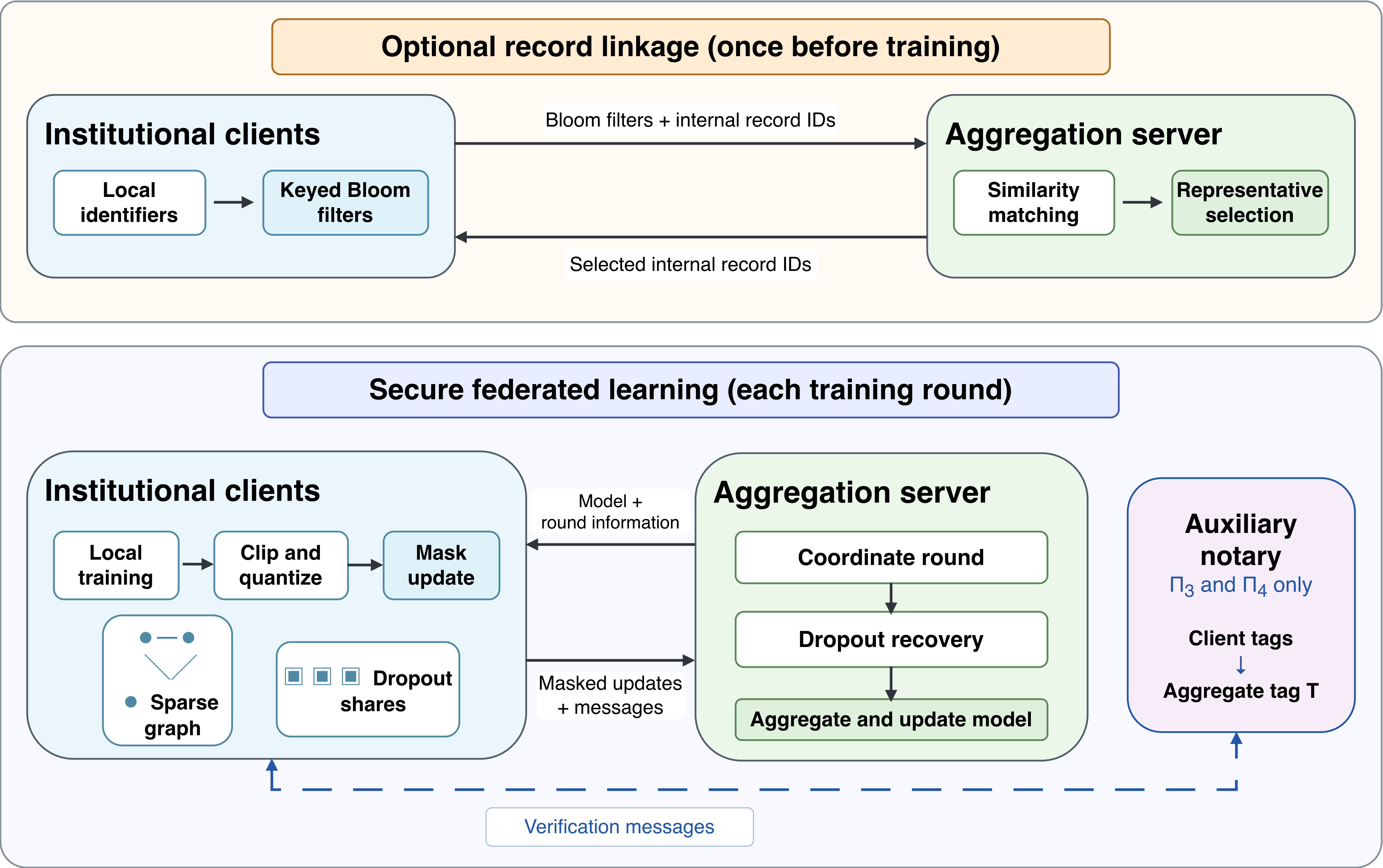}
  \caption{High-level architecture of the proposed framework.}
  \label{fig:framework-architecture}
\end{figure}

Figure~\ref{fig:framework-architecture} presents a high-level view of the proposed framework, showing the data flows during the pre-training record-linkage initialization and during each secure federated learning round. For record linkage, each institutional client constructs Bloom filters from its local identifiers using HMAC with the shared secret key $K$ and sends the resulting Bloom filters, together with internal record identifiers, to the aggregation server. The server computes the similarity between the encoded records, identifies potential matches, selects representative records, and returns the selected internal record identifiers to the corresponding
clients. Plaintext identifiers are not transmitted to the server.  
During each training round, clients train locally, clip and quantize their updates, and protect them using a self-mask and pairwise masks established with neighbors in the sparse communication graph. Secret-shared masking material supports dropout recovery when required. The aggregation server coordinates the round, aggregates the protected updates from the surviving clients, and updates the global model. In variants $\Pi_3$ and $\Pi_4$, the auxiliary notary publishes the public round seed, receives verification tags from the clients, and broadcasts the corresponding aggregate tag used to check the aggregate released by the server. Raw EEG recordings remain within the institutional clients throughout the process.

\subsection{Semi-Honest Secure Aggregation}
The semi-honest protocol follows a masking-based secure aggregation design in which the server coordinates the round but should learn only the aggregate of the surviving client updates. The server is assumed to follow the protocol, but it is not trusted with individual plaintext updates. Each client therefore submits a masked version of its clipped local update. The masks are constructed so that they cancel in the aggregate or can be removed through dropout recovery.

At the beginning of each secure aggregation round, each selected client $i \in S_r$ generates two fresh key pairs, $(sk_i^1,pk_i^1)$ and $(sk_i^2,pk_i^2)$. The first key pair is used for pairwise mask generation, while the second key pair is used for encrypted share distribution. The server constructs the round communication graph and informs each client of its neighbor set $N_G(i)$. For every neighbor $j \in N_G(i)$, client $i$ receives the public keys of $j$ and derives pairwise shared randomness using key agreement. The shared secret is then hashed and expanded using a pseudorandom generator to obtain a mask vector in the secure-aggregation domain:
\begin{equation}
a_{i,j}=F\bigl(H(\mathrm{KA}(sk_i^1,pk_j^1))\bigr)\in\mathbb{Z}_M^\ell
\end{equation}
Here, $F$ denotes the PRG expansion into $\ell$ coordinates modulo $M$. The same PRG expansion is used for the self-mask, so $F(b_i)\in\mathbb{Z}_M^\ell$.

Before masking, the effective client update is mapped to a bounded integer secure-aggregation representation. Let $R_Q=2^{22}$ denote the quantization range and $w_{\max}$ the maximum aggregation weight. For client $i$ with $n_i$ training examples, the corresponding encoded weight is $q_i=\mathrm{round}((n_i/w_{\max})R_Q)$, with $d_i=q_i/R_Q$. The update is weighted by $d_i$, clipped, and quantized so that each model coordinate is represented by an integer in $\{0,\ldots,R_Q-1\}$. The encoded aggregation factor $q_i$ is combined with the quantized model coordinates, and the resulting integer secure-aggregation payload is denoted by $z_i=Q(\hat{x}_i,n_i)$. Thus, negative real-valued coordinates are mapped to non-negative quantization indices rather than signed residues modulo $M$.

The masked upload is
\begin{equation}
y_i = z_i + \mathrm{F}(b_i) -
\sum_{\substack{j \in A_1 \cap N_G(i) \\ 0 < j < i}} a_{i,j} +
\sum_{\substack{j \in A_1 \cap N_G(i) \\ i < j \le n}} a_{i,j}
\pmod{M}
\end{equation}
All terms have the same array structure and are combined modulo $M$. After mask removal, the server obtains
\begin{equation}
Y_Q=\sum_{i\in A'_2} z_i \pmod{M}
\end{equation}

For the reported configurations, $n_i\leq w_{\max}$, so $q_i\leq R_Q$. Hence every coordinate of the factor-combined payload is at most $R_Q$, and $|A'_2|R_Q\leq70\cdot2^{22}<2^{32}=M$. Here, factor-combined means that $q_i$ and the quantized model coordinates are encoded together in $z_i$. Therefore, the reconstructed aggregate does not wrap modulo $M$ in the reported experiments.

The protocol uses $M=2^{32}$ for secure aggregation and $q=2^{61}-1$ for AN verification. AN tags are computed directly over $z_i$ before masking. Since no wraparound occurs in the reported settings, the ordinary integer sum is preserved in $Y_Q$, and verification is performed directly as
\begin{equation}
\left\langle u^{(r,\nu)},Y_Q\right\rangle
\stackrel{?}{=}
T^{(r,\nu)}
\pmod q
\end{equation}
No centered signed decoding is applied before this check. After unmasking and, when applicable, AN verification, the accepted integer aggregate is decoded as $Y=Q^{-1}(Y_Q)$. The decoding step recovers the aggregation weights encoded in $z_i$, so $Y$ denotes the resulting weighted aggregate and requires no additional normalization by $|A'_2|$. Here, $A_1$ denotes the clients completing share distribution. Pairwise masks use opposite signs at their endpoints and therefore cancel when both endpoints contribute.

Client dropout is handled through threshold secret sharing and encrypted share distribution. Each client creates Shamir secret shares of both its self-mask seed $b_i$ and its first private key $sk_i^1$:
\begin{equation}
H_i^b=\mathrm{ShamirSS}(t,k,b_i), \qquad
H_i^s=\mathrm{ShamirSS}(t,k,sk_i^1)
\end{equation}
where $t$ is the reconstruction threshold and $k$ is the number of communication neighbors. For each neighbor $j \in N_G(i)$, client $i$ derives an encryption key using the second key pair, $k_{i,j}=\mathrm{KA}(sk_i^2,pk_j^2)$ and sends the corresponding shares through authenticated encryption:
\begin{equation}
c_{i,j}=\mathrm{Eauth.Enc}\bigl(k_{i,j};\, i \parallel j \parallel h_{i,j}^{b} \parallel h_{i,j}^{s}\bigr)
\end{equation}
The ciphertexts are relayed by the server to the appropriate neighbors. Authenticated encryption protects the confidentiality and integrity of the transmitted shares, while the threshold structure allows reconstruction only when enough neighboring clients provide valid shares.

After the masked-update upload phase, the server identifies the set $A'_2$ of clients whose masked updates were received on time. For every surviving client $i \in A'_2$, the server requests enough shares of $b_i$ from its neighbors, reconstructs the self-mask seed, regenerates $\mathrm{F}(b_i)$, and removes the self-mask from the aggregate. For a client $j$ that dropped out before submitting its masked update, pairwise masks involving $j$ may still affect the masks of surviving neighbors. In this case, the server reconstructs $sk_j^1$ using shares provided by the neighbors of the dropped client, recomputes the corresponding pairwise masks, and removes the residual masks from the aggregate.

The server reconstructs either the self-mask seed or the private masking key of a client, but not both in the same round. This separation is essential because reconstructing both values for a surviving client could reveal enough masking material to expose its plaintext update. After all required self-masks and residual pairwise masks are removed, the server obtains the quantized integer aggregate of the surviving clients:
\begin{equation}
Y_Q = \sum_{i \in A'_2} z_i \pmod{M}
\end{equation}
After optional AN verification, decoding yields the weighted real-valued aggregate $Y=Q^{-1}(Y_Q)$. 
The weighted aggregate is then applied to the global model:
\begin{equation}
w^{(r+1)} = w^{(r)} + \eta Y
\end{equation}
where $\eta$ is the global learning rate. The detailed protocol flow for the semi-honest variants \(\Pi_1\) and \(\Pi_3\) is given in Table~\ref{tab:app-pi1-pi3}.

\begin{table}[!t]
\centering
\caption{Detailed protocol flow for the proposed semi-honest variants $\Pi_1$ and $\Pi_3$. Black rows show steps common to both variants, while red rows show the auxiliary-notary-based verifiability steps specific to $\Pi_3$.}
\label{tab:app-pi1-pi3}
\fontsize{10pt}{10pt}\selectfont
\setlength{\tabcolsep}{3pt}
\renewcommand{\arraystretch}{1.25}
\newcommand{\tabspace}{\noalign{\vskip 1pt}}
\begin{tabular}{p{0.06\textwidth} p{0.94\textwidth}}
\hline
\textbf{Step} & \textbf{Operation} \\
\hline
\tabspace
0 & Define $A_1,A_2,A'_2,A_3$ as clients active at successive stages, where $A_1$ completes share distribution, $A_2$ uploads masked inputs, and $A_3$ responds in recovery; timely subsets $A'_i$ satisfy $A_1\supseteq A'_1\supseteq A_2\supseteq A'_2\supseteq A_3\supseteq A'_3$. Define vector length $\ell$, AN verification-check index $\nu=1,\dots,v$, secure-aggregation modulus $M=2^{32}$, verification modulus $q=2^{61}-1$, quantization range $R_Q=2^{22}$, and PRG $F:\{0,1\}^{\lambda}\rightarrow\mathbb{Z}_M^\ell$, so that $z_i$, $F(b_i)$, $a_{i,j}$, and $y_i$ are all represented in $\mathbb{Z}_M^\ell$. \\
\hline
\tabspace
1 & Server broadcasts current global model $w^{(r)}$; each selected client trains locally on private data and computes update vector $x_i$. \\
\textcolor{red}{1a} & \textcolor{red}{AN publishes public randomness seed $\rho_r$; all clients derive verification vectors $u^{(r,\nu)}=\mathrm{PRG}(\rho_r,\nu)$ for $\nu=1,\dots,v$.} \\
\hline
\tabspace
2 & Server generates a $k$-regular undirected communication graph $G$ using graph procedure $\mathcal{G}_1$ and sends each client its neighbor set $N_G(i)$. \\
\hline
\tabspace
3 & Each client $i$ generates two key pairs $(sk_i^1,pk_i^1)$ and $(sk_i^2,pk_i^2)$; public keys are sent to the server and forwarded to all neighbors $j\in N_G(i)$. \\
\hline
\tabspace
4 & Each client samples $b_i$, computes $H_i^b=\mathrm{ShamirSS}(t,k,b_i)$ and $H_i^s=\mathrm{ShamirSS}(t,k,sk_i^1)$, derives $k_{i,j}=KA(sk_i^2,pk_j^2)$ for each $j\in N_G(i)$, encrypts $c_{i,j}=E_{\mathrm{auth}}.\mathrm{Enc}(k_{i,j},i\Vert j\Vert h^b_{i,j}\Vert h^s_{i,j})$, and sends $m=(j,c_{i,j})$ to the server; clients completing this stage form $A_1$, and server aborts if $|A'_1|<(1-\delta)n$. \\
\hline
\tabspace
5 & Server forwards all ciphertexts $c_{i,j}$ to recipient client $j$; each client decrypts received ciphertexts and determines which neighbors belong to $A'_1$. \\
\hline
\tabspace
6 & Each client computes pairwise masks $a_{i,j}=F(H(\mathrm{KA}(sk_i^1,pk_j^1)))$ for $j\in N_G(i)$ and self-mask $F(b_i)$; then applies $\ell_2$ clipping:
$\tilde{x}_i=x_i$ if $\lVert x_i\rVert_2\le\tau$, and $\tilde{x}_i=\frac{\tau}{\lVert x_i\rVert_2}x_i$ otherwise. \\
6a & Optional future privacy step: set $\hat{x}_i=\mathrm{Anonymize}(\tilde{x}_i)$; in the reported experiments, this was disabled and $\hat{x}_i=\tilde{x}_i$. The effective update and its aggregation weight are encoded into the integer secure-aggregation payload $z_i=Q(\hat{x}_i,n_i)$. \\
\textcolor{red}{6b} & \textcolor{red}{Client computes verification tags, $tag_i^{(r,\nu)}=\langle u^{(r,\nu)},z_i\rangle\bmod q$, directly over the factor-combined integer payload and sends the tags
to the AN.} \\
\hline
\tabspace
7 & Client uploads masked integer vector $y_i=z_i+F(b_i)-\sum_{\substack{j\in A_1\cap N_G(i)\\0<j<i}}a_{i,j}+\sum_{\substack{j\in A_1\cap N_G(i)\\i<j\le n}}a_{i,j}\pmod{M}$; clients uploading masked inputs form $A_2$, and server aborts if $|A'_2|<(1-\delta)n$. \\
\hline
\tabspace
8 & For each client $i\in A'_2$, server sends recovery sets $R_1=A'_2\cap N_G(i)$ and $R_2=(A_1\setminus A'_2)\cap N_G(i)$. \\
\hline
\tabspace
9 & Each client $j$ receives $R_1,R_2$, decrypts stored ciphertexts, and sends $\{(i,h^b_{i,j})\}_{i\in R_1}\cup\{(i,h^s_{i,j})\}_{i\in R_2}$ to the server. \\
\hline
\tabspace
10 & Server collects shares from responders and aborts if $|A'_3|<(1-\delta)n$. \\
\textcolor{red}{10a} & \textcolor{red}{The server sends the finalized contributor set $A'_2$ to the AN, which aggregates the corresponding verification tags as $T^{(r,\nu)}=\sum_{i\in A'_2}tag_i^{(r,\nu)}\bmod q$.} \\
\hline
\tabspace
11 & For each $i\in A'_2$, collect $B_i\subseteq H_i^b$, abort if $|B_i|<t$, reconstruct $b_i$, compute $F(b_i)$, and remove self-mask; for each $i\in A_1\setminus A'_2$, collect $Set_i\subseteq H_i^s$, abort if $|Set_i|<t$, reconstruct $sk_i^1$, and remove corresponding pairwise masks. After removing all self-masks and residual pairwise masks, the server reconstructs the quantized integer aggregate $Y_Q=\sum_{i\in A'_2}z_i\pmod{M}$. \\
\textcolor{red}{11a} & \textcolor{red}{The server broadcasts $Y_Q$ to clients and the AN broadcasts $T^{(r,\nu)}$; clients verify $\langle u^{(r,\nu)},Y_Q\rangle\stackrel{?}{=}T^{(r,\nu)}\pmod q$ for all $\nu=1,\dots,v$, rejecting the round if any check fails.} \\
\hline
\tabspace
12 & After reconstruction, and after successful AN verification when used, the accepted integer aggregate is decoded as $Y=Q^{-1}(Y_Q)$, yielding the weighted aggregate. The server updates $w^{(r+1)}=w^{(r)}+\eta Y$ and broadcasts the new model. \\
\hline
\tabspace
\end{tabular}
\end{table}

\subsection{Malicious Secure Aggregation Extension}
The malicious-setting extension strengthens the semi-honest protocol against server-side inconsistencies by adding commitment, proof verification, signed inclusion evidence, and signed acknowledgment evidence. The protocol tracks client progress through nested sets $[n] \supseteq A_1 \supseteq A'_1 \supseteq A_2 \supseteq A'_2 \supseteq A_3 \supseteq A'_3 \supseteq A_4 \supseteq A'_4$ where each $A_i$ denotes the clients that successfully send messages in a protocol stage, and each $A'_i$ denotes the subset whose messages arrive at the server in time. This ordering ensures that clients can only drop out as the round progresses, and the final aggregation is computed over $A'_2$. The malicious-setting protocol assumes an honest setup phase for initial client key generation, public-key registration, and neighbor-set submission. This does not mean that the server is trusted during the rest of the protocol. Rather, after the setup material is fixed, the server commits to the public-key vectors using a Merkle tree, and clients verify Merkle proofs before accepting neighbor keys. The malicious-setting mechanisms therefore apply after registration, using commitments, proof checks, signed inclusion messages, and acknowledgments to constrain later server behavior and active-set/recovery decisions. In the malicious variant, $p$ denotes the acknowledgment threshold: a client releases recovery shares only after receiving at least $p$ valid acknowledgment signatures from its outgoing neighbors. This parameter is chosen consistently with the reconstruction threshold, and in the experiments we set $p=t$. The table also uses the bound $3k+k=4k$ for the maximum number of public keys a client accepts, where $k$ corresponds to outgoing neighbors and $3k$ is the allowed upper bound for incoming-neighbor keys in the directed sparse graph.

The first difference from the semi-honest setting is the commitment to public keys. Each client generates two key pairs, $K^1_i=(sk^1_i,pk^1_i),  K^2_i=(sk^2_i,pk^2_i)$ where the first key pair is used for pairwise masking and the second key pair is used for authenticated encryption and signatures. The server commits to the public-key vectors $pk^1=(pk^1_i)_i$, $pk^2=(pk^2_i)_i$ using a Merkle tree. When neighbor keys are distributed later, the server attaches Merkle proofs. Clients verify these proofs before accepting the received neighbor keys, and abort if a proof fails. This binds the round to a single public-key view and reduces the ability of the server to present inconsistent key material to different clients. The second difference is the use of a sparse directed communication graph. Each client independently samples $k$ outgoing neighbors $N_{out}(i)$, while the server computes the corresponding incoming sets $N_{in}(i)$ and distributes $\bigl(N_{in}(i),(pk_j^1,pk_j^2)_{j\in N(i)}\bigr)$ with the required Merkle proofs. Clients check that the number of received keys does not exceed the allowed bound, that all Merkle proofs are valid, and that all required neighbor keys are present. The third difference is signed participation evidence. After computing the masked update, client $i$ signs inclusion messages for its active neighbors:
\begin{equation}
m_{i,j}=(\text{"included"}\Vert i\Vert j),
\qquad
\sigma^{\mathrm{incl}}_{i,j}=\mathrm{Sign}(sk^2_i,m_{i,j})
\end{equation}
The client sends $\big(y_i,(m_{i,j},\sigma^{\mathrm{incl}}_{i,j})_{j\in A_{2,i}}\big)$ to the server. The server forwards the inclusion signatures to the relevant clients, who verify signature validity, check that no client appears as both active and dropped, and abort on inconsistency.

The fourth difference is acknowledgment-based control of recovery-share release. For each active neighbor $j$, client $i$ signs $m_{i,j}=(\text{"ack"}\Vert i\Vert j)$ to produce $\sigma^{\mathrm{ack}}_{i,j}$. A client proceeds to reveal the corresponding Shamir's shares only after receiving enough valid acknowledgments from its outgoing neighbors. This prevents the server from fabricating inclusion claims and helps ensure that recovery material is released only under a consistent participation view. The server then reconstructs $b_i$ for $i\in A'_2$ to remove self-masks, and reconstructs the required masking material for clients in $A_1\setminus A'_2$ to remove residual pairwise masks of dropped clients. The remaining aggregate is then computed as before.

\subsection{Auxiliary-Notary Verifiability}
The verifiability extension provides lightweight checking that the released aggregate is consistent with the verification tags associated with the declared contributor set. The auxiliary notary (AN) provides a lightweight consistency-checking layer inspired by Linear Verification Sketch Aggregation \cite{LiZD25}. The AN does not receive raw EEG data or plaintext model updates, and it is assumed not to collude with the aggregation server. However, it receives low-dimensional linear verification tags computed from client updates, which may reveal limited linear information about the protected update vectors. The AN aggregates these tags so that clients can check whether the server's released aggregate is consistent with the tags submitted for the active set $A'_2$. In this implementation, the verification randomness is public before the server releases the aggregate; therefore, the AN mechanism is interpreted as lightweight aggregate-consistency checking rather than full adaptive malicious-output protection.

At the beginning of round $r$, the AN publishes a public randomness seed $\rho_r$. All clients use this seed to derive $v$ public verification vectors, indexed by $\nu=1,\dots,v$:
\begin{equation}
u^{(r,\nu)}=\mathrm{PRG}(\rho_r,\nu),
\qquad \nu=1,\dots,v
\end{equation}
The parameter $v$ controls the number of independent verification checks. Each client first clips its local update $x_i$ to obtain $\tilde{x}_i$. If an additional optional privacy transformation is enabled, the clipped update is transformed into the effective submitted update $\hat{x}_i=\mathrm{Anonymize}(\tilde{x}_i)$. In the reported experiments, no additional anonymization method was enabled and, therefore, $\hat{x}_i=\tilde{x}_i$. This step is separate from Bloom filter-based record linkage, which is a pre-training identifier-linkage component, and from secure aggregation, which protects model updates during training. The effective update and its aggregation weight are then encoded into the bounded integer secure-aggregation representation defined above, denoted by $z_i=Q(\hat{x}_i,n_i)$. For each verification vector, client $i$ computes:
\begin{equation}
tag_i^{(r,\nu)} = 
\left\langle u^{(r,\nu)},z_i\right\rangle
\bmod q,
\qquad \nu=1,\dots,v 
\end{equation}
The masked integer update is sent to the server, while the verification tags are sent to the AN. After secure aggregation, the server reconstructs the integer aggregate
\begin{equation}
Y_Q = 
\sum_{i\in A'_2} z_i
\pmod{M}
\end{equation}
After masked-update collection, the server sends the finalized contributor set $A'_2$ to the AN, which uses it to select the corresponding previously submitted tags and computes:
\begin{equation}
T^{(r,\nu)} = 
\sum_{i\in A'_2} tag_i^{(r,\nu)}
\bmod q 
\end{equation}

Since the AN obtains $A'_2$ from the server, the set is not independently validated. Thus, verification establishes consistency  relative to the server-declared contributor set, but does not detect consistent omission from both $A'_2$ and the aggregate. Verification is performed directly on the reconstructed integer aggregate before factor extraction and dequantization. Since the reported configurations do not wrap modulo $M$, each client verifies
\begin{equation}
\left\langle u^{(r,\nu)},Y_Q\right\rangle
\stackrel{?}{=}
T^{(r,\nu)}
\pmod q,
\qquad \forall \nu=1,\dots,v 
\end{equation}
If all checks are satisfied, the aggregate is accepted as consistent with the AN-aggregated tags. Decoding $Y=Q^{-1}(Y_Q)$ then yields the weighted aggregate used for the global model update. If any check fails, the round is rejected because the released aggregate is inconsistent with the AN-aggregated tags. Since the verification vectors are public before the server releases the aggregate, this check provides lightweight consistency evidence, but it is not a full commit-then-challenge proof against an adaptively chosen malicious output. The same verifiability mechanism can be applied to both the semi-honest and malicious secure aggregation variants without changing their masking structure. 
The detailed protocol flow for the malicious-setting variants \(\Pi_2\) and \(\Pi_4\) is given in Table~\ref{tab:app-pi2-pi4}.

\begin{table}[!t]
\centering
\vspace{-10pt}
\caption{Detailed protocol flow for the proposed malicious-setting variants $\Pi_2$ and $\Pi_4$. Black rows show steps common to both variants, while red rows show the auxiliary-notary-based verifiability steps specific to $\Pi_4$.}
\label{tab:app-pi2-pi4}
\fontsize{9.5pt}{9pt}\selectfont
\setlength{\tabcolsep}{2pt}
\renewcommand{\arraystretch}{1.16}
\newcommand{\tabspace}{\noalign{\vskip 1pt}}
\begin{tabular}{p{0.06\textwidth} p{0.94\textwidth}}
\hline
\textbf{Step} & \textbf{Operation} \\
\hline
\tabspace
0 & Define $A_1,A_2,A_3,A_4$ as the sets of clients that send the required messages at successive stages of the malicious protocol. For $i=1,\ldots,4$, let $A'_i\subseteq A_i$ denote the subset of those clients whose messages reach the server on time. The sets satisfy $[n]\supseteq A_1\supseteq A'_1\supseteq A_2\supseteq A'_2
\supseteq A_3\supseteq A'_3\supseteq A_4\supseteq A'_4$, and the final aggregate is computed over $A'_2$. Define vector length $\ell$, AN verification-check index $\nu=1,\dots,v$, secure-aggregation modulus $M=2^{32}$, verification modulus $q=2^{61}-1$, quantization range $R_Q=2^{22}$, and PRG $F:\{0,1\}^{\lambda}\rightarrow\mathbb{Z}_M^\ell$, so that the quantized update $z_i$, self-mask $F(b_i)$, pairwise masks $a_{i,j}$, and masked upload $y_i$ are all represented in $\mathbb{Z}_M^\ell$. \\
\hline
\tabspace
1 & During the honest setup phase for initial key generation and registration, each client $i$ generates $K_i^1=(sk_i^1,pk_i^1)$ and $K_i^2=(sk_i^2,pk_i^2)$, then sends $(pk_i^1,pk_i^2)$ to the server. \\
\hline
\tabspace
2 & After setup registration, the server commits to public-key vectors $pk^1=(pk_i^1)_i$ and $pk^2=(pk_i^2)_i$ using a Merkle tree. \\
\textcolor{red}{2a} & \textcolor{red}{For the AN variant, the AN publishes public seed $\rho_r$; clients derive $u^{(r,\nu)}=\mathrm{PRG}(\rho_r,\nu)$ for $\nu=1,\dots,v$.} \\
\hline
\tabspace
3 & Clients and server jointly generate directed graph $G([n],E)$: each client $i$ samples $k$ outgoing neighbors $N_{out}(i)$ without replacement from $[n]$ and sends $N_{out}(i)$ to the server; $N_{in}(i)=\{j:i\in N_{out}(j)\}$ and $N(i)=N_{in}(i)\cup N_{out}(i)$. \\
\hline
\tabspace
4 & Server sends $\big(N_{in}(i),(j,pk_j^1,pk_j^2)_{j\in N(i)}\big)$ to client $i$, together with $|N(i)|\log_2(n)$ Merkle hashes for public-key verification. \\
\hline
\tabspace
5 & Client aborts if the server sends more than $4k$ public keys, corresponding to at most $k$ outgoing-neighbor keys and $3k$ incoming-neighbor keys; it verifies Merkle consistency of received public keys and checks that all public keys for $N_{out}(i)$ are present. \\
\hline
\tabspace
6 & Each client samples $b_i$, computes $H_i^b=\{h^b_{i,1},\dots,h^b_{i,k}\}=\mathrm{ShamirSS}(t,k,b_i)$ and $H_i^s=\{h^s_{i,1},\dots,h^s_{i,k}\}=\mathrm{ShamirSS}(t,k,sk_i^1)$; for each $j\in N_{out}(i)$, it sends $m_j=(j,c_{i,j})$ where $c_{i,j}=E_{\mathrm{auth}}.\mathrm{Enc}(k_{i,j},i\Vert j\Vert h^b_{i,j}\Vert h^s_{i,j})$ and $k_{i,j}=KA(sk_i^2,pk_j^2)$. \\
\hline
\tabspace
7 & Server aborts if $|A'_1|<(1-\delta)n$; otherwise, it forwards all messages $(j,c_{i,j})$ to client $j$, defining $A_{2,j}\subseteq N(j)$ as the clients from which $j$ received ciphertexts. \\
\hline
\tabspace
8 & Each client decrypts received ciphertexts and aborts on decryption failure; computes $a_{i,j}=F(H(\mathrm{KA}(sk_i^1,pk_j^1)))$ for $j\in A_{2,i}$ and self-mask $F(b_i)$; clips $x_i$ to $\tilde{x}_i$, where $\tilde{x}_i=x_i$ if $\lVert x_i\rVert_2\le\tau$, and $\tilde{x}_i=\frac{\tau}{\lVert x_i\rVert_2}x_i$ otherwise. \\
8a & Optional future privacy step: set $\hat{x}_i=\mathrm{Anonymize}(\tilde{x}_i)$; disabled in the reported experiments, so $\hat{x}_i=\tilde{x}_i$. The effective update and its aggregation weight are encoded into the integer secure-aggregation payload $z_i=Q(\hat{x}_i,n_i)$. \\
\textcolor{red}{8b} & \textcolor{red}{For the AN variant, client computes verification tags as $tag_i^{(r,\nu)}=\langle u^{(r,\nu)},z_i\rangle\bmod q$ directly over the factor-combined integer payload and sends the tags to the AN.} \\
\hline
\tabspace
9 & Client computes the masked integer input $y_i=z_i+F(b_i)-\sum_{j\in A_{2,i},\,0<j<i}a_{i,j}+\sum_{j\in A_{2,i},\,i<j\le n}a_{i,j}\pmod{M}$, signs $m_{i,j}=(\text{"included"}\Vert i\Vert j)$ using $sk_i^2$ to obtain $\sigma^{\mathrm{incl}}_{i,j}$ for all $j\in A_{2,i}$, and sends $\big(y_i,(m_{i,j},\sigma^{\mathrm{incl}}_{i,j})_{j\in A_{2,i}}\big)$ to the server. \\
\hline
\tabspace
10 & Server aborts if $|A'_2|<(1-\delta)n$; otherwise, for each $i\in A'_2$, it sends $(A'_2\cap N_{in}(i),(A_1\setminus A'_2)\cap N_{in}(i))$ and all inclusion messages/signatures $(m_{j,i},\sigma^{\mathrm{incl}}_{j,i})$, defining $A^b_{3,i}$ and $A^s_{3,i}$. \\
\hline
\tabspace
11 & Each client checks $A^b_{3,i}\cap A^s_{3,i}=\emptyset$, $A^b_{3,i},A^s_{3,i}\subseteq N_{in}(i)\cap A_{2,i}$, and verifies all $\sigma^{\mathrm{incl}}_{j,i}$ for $j\in A^b_{3,i}$; it aborts on failure. \\
\hline
\tabspace
12 & For every $j\in A^b_{3,i}\subseteq N_{in}(i)$, client $i$ signs $m_{i,j}=(\text{``ack''}\Vert i\Vert j)$ using $sk_i^2$ to obtain $\sigma^{\mathrm{ack}}_{i,j}$ and sends it to the server. \\
\hline
\tabspace
13 & Server aborts if $|A'_3|<(1-\delta)n$; otherwise, it forwards all acknowledgment messages and signatures to client $j$. \\
\hline
\tabspace
14 & Each client verifies received acknowledgment signatures using public keys and aborts on failure; once client $j$ obtains at least $p$ valid signatures from neighbors in $N_{out}(j)$, it sends $\{(i,h^b_{i,j})\}_{i\in A^b_{3,j}}\cup\{(i,h^s_{i,j})\}_{i\in A^s_{3,j}}$ to the server. \\
\hline
\tabspace
15 & Server aborts if $|A'_4|<(1-\delta)n$. \\
\textcolor{red}{15a} & \textcolor{red}{For the AN variant, the server sends the finalized contributor set $A'_2$ to the AN, which computes $T^{(r,\nu)}=\sum_{i\in A'_2}tag_i^{(r,\nu)}\bmod q$.} \\
\hline
\tabspace
16 & For each $i\in A'_2$, it collects $B_i\subseteq H_i^b$, aborts if $|B_i|<t$, reconstructs $b_i$, and removes $F(b_i)$; for each $i\in A_1\setminus A'_2$, it collects $Set_i\subseteq H_i^s$, aborts if $|Set_i|<t$, reconstructs $sk_i^1$, and removes corresponding pairwise masks. After removing all self-masks and residual pairwise masks, the server reconstructs $Y_Q=\sum_{i\in A'_2}z_i\pmod M$. For the base malicious variant $\Pi_2$, decoding $Y=Q^{-1}(Y_Q)$ yields the weighted aggregate used for the global update. \\
\textcolor{red}{16a} & \textcolor{red}{For the AN variant, the server broadcasts $Y_Q$ to clients and the AN broadcasts $T^{(r,\nu)}$; clients verify $\langle u^{(r,\nu)},Y_Q\rangle\stackrel{?}{=}T^{(r,\nu)}\pmod q$ for all $\nu=1,\dots,v$, rejecting the round if any check fails. After successful verification, decoding $Y=Q^{-1}(Y_Q)$ yields the weighted aggregate used for the global update.} \\
\hline
\tabspace
\end{tabular}
\end{table}

\subsection{Privacy-Preserving Record-Linkage Initialization}
The Bloom filter-based record-linkage step is an initialization/pre-training component used before federated learning begins, rather than a component executed during secure aggregation rounds. All clients agree on the Bloom filter length $len$, the q-gram length $q_g$, number of hash functions $d$, matching threshold $\gamma$, and a shared secret key $K$ used only for local identifier encoding. For each identifier string $str$, the client pads the string and extracts the multiset of q-grams $\mathrm{Frag}_{q_g}(str)$. A Bloom filter $\mathrm{BF}\in\{0,1\}^{len}$ is initialized to zero. For each fragment $g\in\mathrm{Frag}_{q_g}(str)$ and each hash index $e\in\{1,\dots,d\}$, the client computes
\begin{equation}
h_e=\mathrm{HMAC}_K(g\parallel e),
\qquad
\mathrm{index}_e=h_e \bmod len 
\end{equation}
and sets $\mathrm{BF}[\mathrm{index}_e]\leftarrow 1$. The plaintext identifier is kept locally and is not sent to the server. The server receives only Bloom filter representations and internal record identifiers. For two Bloom filters, it computes the Dice--S{\o}rensen similarity coefficient
\begin{equation}
\mathrm{DSC} = \frac{2\mathrm{TP}}{2\mathrm{TP}+\mathrm{FP}+\mathrm{FN}}
\end{equation}
where $\mathrm{TP}$ is the number of bit positions set to one in both filters, $\mathrm{FP}$ is the number of positions where only the second filter is one, and $\mathrm{FN}$ is the number of positions where only the first filter is one. Two records are treated as a potential match if $\mathrm{DSC}\ge \gamma$. Matched internal identifiers are grouped into equivalence classes, and a deterministic representative is selected so that at most one record per linked individual is used for federated training. This component handles duplicate resolution during data preparation, while secure aggregation protects model updates during training \cite{SchnellBR09,VatsalanSCR17}. When linkage attributes are unavailable or have already been removed, the initialization step is bypassed without changing the subsequent secure aggregation protocol. Because the current dataset contained no usable linkage attributes, empirical evaluation of linkage quality and module overhead is left for future work.
Figures~\ref{fig:pi3-flow} and~\ref{fig:pi4-flow} summarize the end-to-end message flows of the semi-honest and malicious secure aggregation variants, respectively, including the optional auxiliary-notary verification steps.

\clearpage

\begin{landscape}
\begingroup
\centering
\vspace{-3cm}
\tiny
\setlength{\tabcolsep}{2.9pt}

\scalebox{0.98}{%
\makebox[\linewidth][c]{%
\begin{tabular}{@{}c c c c c c@{}}
\textcolor{red}{\textbf{\underline{AN}}} & & \textbf{\underline{Client $i$}} & & \textbf{\underline{Server}} & \\[2mm]
\hypertarget{protocol-pi3}{}

& & Bloom Filter

& $\overset{\textbf{0: }\mathrm{BF}_i,\mathrm{id}_i}{\xrightarrow{\hspace*{1.5cm}}}$ & Bloom Filter \\

& & Generate $(sk_i^1,pk_i^1),(sk_i^2,pk_i^2)$ & & & \\

& & & $\overset{\textbf{0: }pk_i^1,pk_i^2}{\xrightarrow{\hspace*{1.5cm}}}$ & Collect public keys & \\

& & & $\overset{\textbf{1: }w^{(r)}}{\xleftarrow{\hspace*{1.5cm}}}$ & & \\

& & Compute local update $x_i^{(r)}$ & & & \\

& & & & Generate graph as before & \\[1mm]

& & & $\overset{\textbf{2: }N_G(i), pk_j^1, pk_j^2}{\xleftarrow{\hspace*{1.5cm}}}$ & & \\

& & Generate $F(b_i)$ & & & \\
& & $H_i^b=\textsf{ShamirSS}(t,k,b_i)$, $H_i^s=\textsf{ShamirSS}(t,k,sk_i^1)$ & & & \\

& & $k_{i,j}=KA(sk_i^2,pk_j^2)$ & & & \\
& & $c_{i,j}=E_{\mathrm{auth}}.\mathrm{Enc}(k_{i,j},i\parallel j\parallel h_{i,j}^b\parallel h_{i,j}^s)$ & & & \\

& & & $\overset{\textbf{3: }m=(j,c_{i,j})}{\xrightarrow{\hspace*{1.5cm}}}$ & Receive encrypted shares & \\

& & & & If $|A'_1|<(1-\delta)n$ abort else forward & \\

& & & $\overset{\textbf{4: }c_{i,j}}{\xleftarrow{\hspace*{1.5cm}}}$ & & \\

& & $a_{i,j}=F(H(KA(sk_i^1,pk_j^1)))\in\mathbb{Z}_M^\ell$ & \\ 
& & $\tilde{x}_i=\min\!(1,\frac{\tau}{\lVert x_i\rVert_2})x_i$ & & & \\

& & $\hat{x}_i=\tilde{x}_i$; \textcolor{red}{optional: $\hat{x}_i=\mathrm{Anonymize}(\tilde{x}_i)$} & \\
& & $z_i=Q(\hat{x}_i,n_i)\in\mathbb{Z}_M^\ell$, where $M=2^{32}$ and $R_Q=2^{22}$ & & & \\

\textcolor{red}{Publish seed} 
& \textcolor{red}{$\overset{\textbf{5.1: }\rho_r}{\xrightarrow{\hspace*{1.5cm}}}$}
& \textcolor{red}{$u^{(r,\nu)}=\mathrm{PRG}(\rho_r,\nu)$, $tag_i^{(r,\nu)}=\langle u^{(r,\nu)},z_i\rangle \bmod q$} \\
& \textcolor{red}{$\overset{\textbf{5.2: }tag_i^{(r,\nu)}}{\xleftarrow{\hspace*{1.5cm}}}$} & & & \\
\textcolor{red}{Collect tags}
& & & \\

& & $y_i=z_i+F(b_i) -\sum\limits_{\substack{j\in A_1\cap N_G(i)\\0<j<i}}a_{i,j}
+\sum\limits_{\substack{j\in A_1\cap N_G(i)\\i<j\le n}}a_{i,j}\pmod{M}$ 
& & & \\

& & & $\overset{\textbf{6: }y_i}{\xrightarrow{\hspace*{1.5cm}}}$ & Collect masks if $|A'_2|<(1-\delta)n$ abort & \\
& & & & $R_1=A'_2\cap N_G(i)$, $R_2=(A_1\setminus A'_2)\cap N_G(i)$ & \\

& & & $\overset{\textbf{7: }R_1,R_2}{\xleftarrow{\hspace*{1.5cm}}}$ & & \\

& & Decrypt and Send $\{(i,h^b_{i,j})\}_{i\in R_1}\cup\{(i,h^s_{i,j})\}_{i\in R_2}$ \\
& & & $\overset{\textbf{8: }shares}{\xrightarrow{\hspace*{1.5cm}}}$ \\

& & & & If $|A'_3|<(1-\delta)n$ abort & \\ [-2mm] 

&
\multicolumn{3}{c}{
\textcolor{red}{
$\overset{\textbf{9.1: } \text{Finalized }  A'_2}
{\xleftarrow{\hspace*{11cm}}}$}}
& \\

\textcolor{red}{Aggregate $T^{(r,\nu)}=\sum_{i\in A'_2} tag_i^{(r,\nu)}\bmod q$} 

& & & & Recover $b_i$, $sk_i^1$, and $a_{i,j}$ & \\[1mm]

& & & & Reconstruct $Y_Q=\sum_{i\in A'_2}z_i\pmod{M}$ & \\

& & & \textcolor{red}{$\overset{\textbf{9.2: } {Y_Q}}{\xleftarrow{\hspace*{1.5cm}}}$} \\

\textcolor{red}{Broadcast $T^{(r,\nu)}$}
& \textcolor{red}{$\overset{\textbf{9.3: }T^{(r,\nu)}}{\xrightarrow{\hspace*{1.5cm}}}$} \\

& & \textcolor{red}{Verify $\langle u^{(r,\nu)},Y_Q\rangle \stackrel{?}{=} T^{(r,\nu)} \pmod q$} \\
\\

& & & & Decode and dequantize $Y=Q^{-1}(Y_Q)$ & \\
& & & & $w^{(r+1)}=w^{(r)}+\eta Y$ & \\
& & & $\overset{\textbf{10: } w^{(r+1)}}{\xleftarrow{\hspace*{1.5cm}}}$ \\
\end{tabular}
}}
\captionsetup{font=captionbetween}
\captionof{figure}{Protocol flow for the semi-honest secure aggregation variants ($\Pi_1, \Pi_3$), without and with auxiliary-notary verifiability. Red entries indicate auxiliary-notary verification steps and optional update-transformation steps. The optional pre-training record-linkage phase is shown at Step 0.}
\label{fig:pi3-flow}
\endgroup
\end{landscape}

\begin{landscape}
\begingroup
\vspace{-5cm}
\centering
\tiny
\setlength{\tabcolsep}{3pt}

\scalebox{0.98}{%
\makebox[\linewidth][c]{%
\begin{tabular}{@{}c c c c c c@{}}
\textcolor{red}{\textbf{\underline{AN}}} & & \textbf{\underline{Client $i$}} & & \textbf{\underline{Server}} & \\[2mm]
\hypertarget{protocol-pi4}{}

& & Bloom Filter 
& $\overset{\textbf{0: }\mathrm{BF}_i,\mathrm{id}_i}{\xrightarrow{\hspace*{1.5cm}}}$ & Bloom Filter \\
& & Generate $(sk_i^1,pk_i^1),(sk_i^2,pk_i^2)$ & & & \\
& & Client samples a set of $k$ outgoing neighbors & & & \\

& & & $\overset{\textbf{1: }pk_i^1,pk_i^2, N_{out}(i)}{\xrightarrow{\hspace*{2cm}}}$ &
Collect keys and build Merkle tree & \\
& & & & Derive $N_{in}(i)$ & \\

& & & $\overset{\textbf{2: }Keys, N_{in}(i), hashes}{\xleftarrow{\hspace*{2cm}}}$ & & \\

& & Verify keys with Merkle and Generate $F(b_i)$ & & & \\
& & $H_i^b=\textsf{ShamirSS}(t,k,b_i)$, $H_i^s=\textsf{ShamirSS}(t,k,sk_i^1)$ & & & \\

& & $k_{i,j}=KA(sk_i^2,pk_j^2)$ & & & \\
& & $c_{i,j}=E_{\mathrm{auth}}.\mathrm{Enc}(k_{i,j},i\parallel j\parallel h_{i,j}^b\parallel h_{i,j}^s)$ & & & \\

& & & $\overset{\textbf{3: }m=(j,c_{i,j})}{\xrightarrow{\hspace*{2cm}}}$ & Collect ciphertext and abort; if $|A'_1|<(1-\delta)n$ & \\

& & & $\overset{\textbf{4: }m = (j, c_{i,j)}}{\xleftarrow{\hspace*{2cm}}}$ & & \\

& & Decrypt and abort if failed & & & \\
& & $a_{i,j}=F(H(KA(sk_i^1,pk_j^1)))\in\mathbb{Z}_M^\ell$ for $j\in A_{2,i}$ & \\
& & $\tilde{x}_i=\min\!(1,\frac{\tau}{\lVert x_i\rVert_2})x_i$, $\hat{x}_i=\tilde{x}_i$; \textcolor{red}{optional: $\hat{x}_i=\mathrm{Anonymize}(\tilde{x}_i)$} & \\
& & $z_i=Q(\hat{x}_i,n_i)\in\mathbb{Z}_M^\ell$, where $M=2^{32}$ and $R_Q=2^{22}$ & & & \\

\textcolor{red}{Publish seed} 
& \textcolor{red}{$\overset{\textbf{4.1: }\rho_r}{\xrightarrow{\hspace*{2cm}}}$}
& \textcolor{red}{$u^{(r,\nu)}=\mathrm{PRG}(\rho_r,\nu)$, $tag_i^{(r,\nu)}=\langle u^{(r,\nu)},z_i\rangle\bmod q$} \\
& \textcolor{red}{$\overset{\textbf{4.2: }tag_i^{(r,\nu)}}{\xleftarrow{\hspace*{2cm}}}$} & & & \\
\textcolor{red}{Collect tags}
& & & \\

& & Calculate $y_i$; $m_{i,j}=(\text{``included''}\Vert i\Vert j)$, $\sigma^{incl}_{i,j}\leftarrow \mathrm{Sign}(sk_i^2,m_{i,j})$ & & & \\

& & & $\overset{\textbf{5: }(y_i,\ (m_{i,j},\sigma^{incl}_{i,j}))}{\xrightarrow{\hspace*{2cm}}}$ & Collect masks and signatures; if $|A'_2|<(1-\delta)n$ abort & \\

& & & $\overset{\textbf{6: }A\text{-sets},(m_{j,i},\sigma^{incl}_{j,i})}{\xleftarrow{\hspace*{2cm}}}$ & & \\

& & Check sets and abort if failed \\
& & $m^{\mathrm{ack}}_{i,j}=(\text{``ack''}\Vert i\Vert j)$, $\sigma^{ack}_{i,j}\leftarrow \mathrm{Sign}(sk_i^2,m^{ack}_{i,j})$
& $\overset{\textbf{7: }(m^{\mathrm{ack}}_{i,j},\sigma^{ack}_{i,j})}{\xrightarrow{\hspace*{2cm}}}$ \\

& & & & Collect messages and aborts if $|A'_3|<(1-\delta)n$ \\

& & & $\overset{\textbf{8: }(j,m_{i,j},\sigma^{ack}_{i,j})}{\xleftarrow{\hspace*{2cm}}}$ \\
& & Collect messages and verify signatures, abort if invalid & \\[1mm]

& & & $\overset{\textbf{9: }\{(i,h^b_{i,j})\}\cup\{(i,h^s_{i,j})\}}{\xrightarrow{\hspace*{2cm}}}$ \\

& & & & Collect all and abort if $|A'_4|<(1-\delta)n$ & \\ [-2mm]
&
\multicolumn{3}{c}{
\textcolor{red}{
$\overset{\textbf{10: } \text{Finalized } A'_2}
{\xleftarrow{\hspace*{11.3cm}}}$}}
& \\

\textcolor{red}{Aggregate $T^{(r,\nu)}=\sum_{i\in A'_2}tag_i^{(r,\nu)}\bmod q$} 

& & & & Reconstruct and recover and output $Y_Q=\sum_{i\in A'_2}z_i\pmod{M}$  & \\

& & & \textcolor{red}{$\overset{\textbf{10.1: } {Y_Q}}{\xleftarrow{\hspace*{2cm}}}$} \\

\textcolor{red}{Broadcast $T^{(r,\nu)}$}

& \textcolor{red}{$\overset{\textbf{10.2: }T^{(r,\nu)}}{\xrightarrow{\hspace*{2cm}}}$} \\

& & \textcolor{red}{Verify $\langle u^{(r,\nu)},Y_Q\rangle \stackrel{?}{=} T^{(r,\nu)}\pmod q$} \\

& & & & Decode and dequantize $Y=Q^{-1}(Y_Q)$ & \\
& & & & $w^{(r+1)}=w^{(r)}+\eta Y$ & \\
& & & $\overset{\textbf{11: } w^{(r+1)}}{\xleftarrow{\hspace*{2cm}}}$ \\
\end{tabular}
}}
\captionsetup{font=captionbetween}

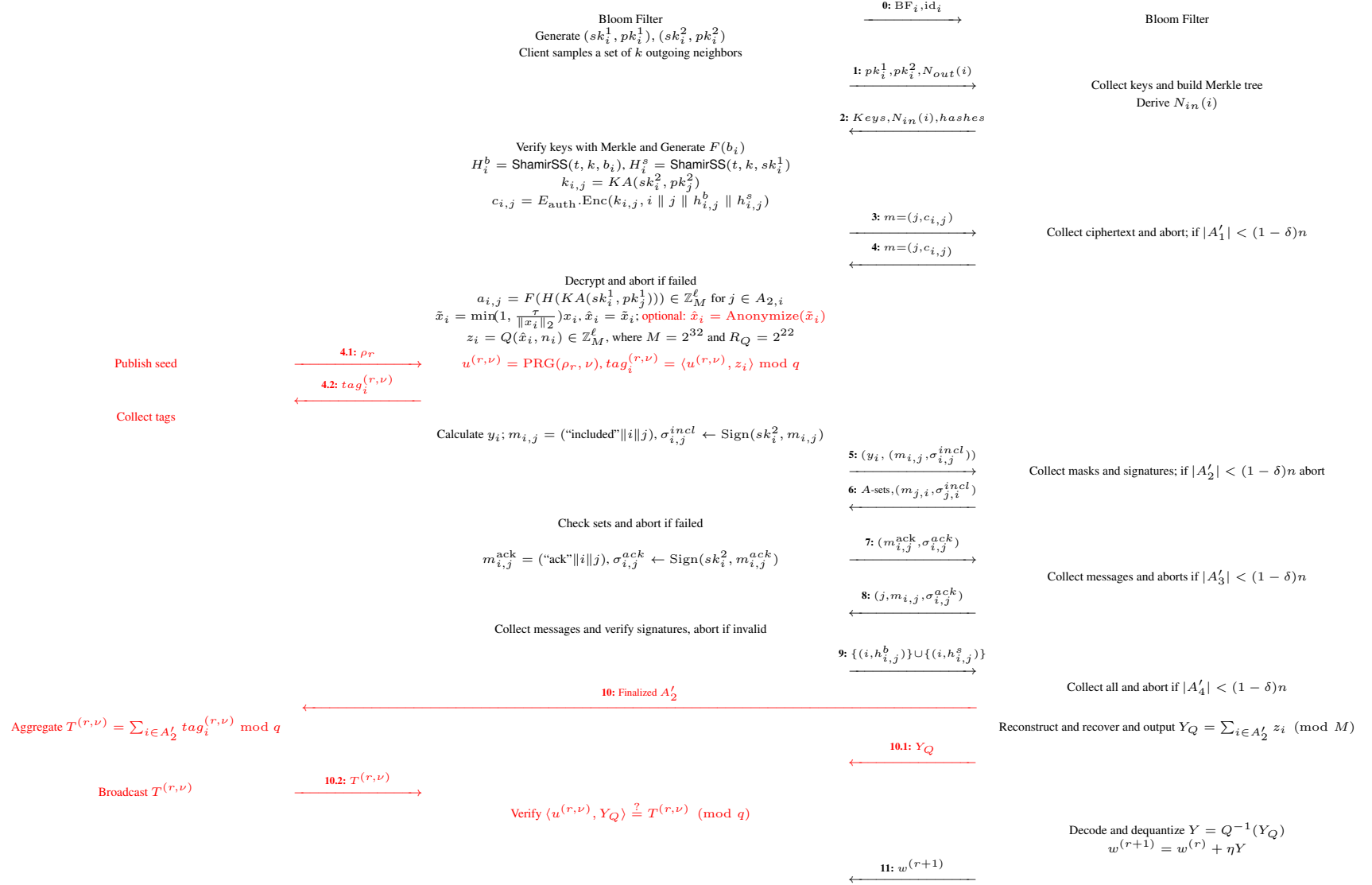
\captionof{figure}{Protocol flow for the malicious-setting aggregation variants ($\Pi_2, \Pi_4$) without and with auxiliary-notary verifiability. Red entries indicate auxiliary-notary verification steps and optional update-transformation steps. The optional pre-training record-linkage phase is shown at Step 0.}
\label{fig:pi4-flow}
\endgroup
\end{landscape}

\section{Experimental Evaluation}
The framework was implemented using the Flower federated learning framework \cite{abs-2007-14390}, which provides a flexible client-server abstraction, supports custom aggregation strategies, and allows controlled simulation of multiple federated clients. The implementation includes a baseline FL pipeline, the basic SecAgg comparison protocol, and the four proposed protocols $\Pi_1$ to $\Pi_4$ described above. Each simulated client performs local training on its assigned EEG data partition and returns either a plaintext update in the baseline setting or a protected update in the secure aggregation settings. The evaluation measures the runtime, communication cost, and learning behavior of the protocol variants under different client configurations.

The experiments use recordings from the TUH Abnormal EEG Corpus (TUAB), version 3.0.1, a subcorpus of the TUH EEG Corpus containing recordings annotated as normal or abnormal \cite{tuh_eeg,obeid2016temple}. The source training pool contained 2,717 EDF recordings, comprising 1,371 normal and 1,346 abnormal recordings. The recordings were organized into the class-specific normal and abnormal directories provided by the dataset and were assigned binary labels 0 and 1, respectively. The source recordings were used to construct configuration-specific client-folder pools. For the 10-client configuration, 10 client folders were created with 250 recordings per folder. For the 40-client configuration, 80 client folders were created with 30 recordings per folder. For the 70-client configuration, 140 client folders were created with 18 recordings per folder. The selected clients in each run were subsequently divided into training and distributed-evaluation clients as described in Table~\ref{tab:experimental_setup}. Each EDF recording was loaded using MNE and resampled to 100Hz. The preprocessing used the signal-channel ordering provided by the source EDF/MNE pipeline and retained the first 19 returned signal channels, with zero-padding when fewer than 19 signals were available. No additional canonical electrode-name remapping was applied. This preprocessing choice limits interpretation of the reported EEG classification accuracy, but does not alter the secure aggregation protocol mechanics or the fixed dimensionality of the model updates used for protocol evaluation. The first 10s after resampling, corresponding to 1,000 temporal samples per channel, were used from each recording, with temporal zero-padding if necessary. Finally, each retained channel was standardized independently over the selected segment by subtracting its mean and dividing by its standard deviation plus $10^{-6}$.
The experiments were executed on a system equipped with a 13th-generation Intel Core i7-13650HX processor, 24~GB of RAM, and an NVIDIA GeForce RTX~5070 GPU. The implementation used Python~3.11.9 and Flower/Ray-based simulation.
The secure-aggregation implementation relied on Flower's secure-aggregation utilities together with NumPy, PyTorch, MNE, and Python cryptographic libraries.
Table~\ref{tab:experimental_setup} summarizes the dataset, preprocessing, model, client partitioning, protocol parameters, and experimental scope. The learning model is a custom lightweight 1D CNN used as a common reference model across all protocol variants. Each EEG input is represented as a $19\times1000$ tensor, where the 19 retained EEG channels form the input channels and the 1,000 samples correspond to the 10s temporal dimension at 100Hz. The network consists of three one dimensional convolutional layers. The first maps 19 input channels to 32 feature maps using a kernel size of 7, stride 1, and padding 3, followed by ReLU activation and max pooling with a factor of 2. The second convolution maps 32 to 64 feature maps with kernel size 7, stride 1, and padding 3, again followed by ReLU activation and max pooling with a factor of 2. The third convolution maps 64 to 128 feature maps using kernel size 5, stride 1, and padding 2, followed by ReLU activation and adaptive average pooling to one temporal value per feature map. The resulting 128-dimensional representation is passed to a fully connected linear layer with two outputs for normal/abnormal classification. The use of 1D convolution provides a lightweight means of extracting temporal features from the multichannel EEG signals. The architecture was not adapted from a specific published EEG model; rather, it was designed as a compact reference classifier for evaluating the secure aggregation protocols. The same CNN architecture and training configuration are used across all protocol variants so that the evaluation compares the secure-aggregation mechanisms under a consistent learning workload rather than comparing EEG classification architectures. Because fixed file-to-client partitions and identical random seeds were not enforced across all protocol variants, runtime comparisons mainly reflect protocol overhead, while accuracy comparisons should be interpreted approximately because they are affected by partition variability and run level randomness. Training and distributed evaluation were separated at the client-folder level. For each run, the selected client folders were divided into approximately 80\% training clients and 20\% distributed evaluation clients, so the training and evaluation clients used different EDF recordings within that run. Across repeated runs, however, client folders could be reused under different training/evaluation assignments as the client selection was rotated. Because the available dataset organization did not provide patient-level grouping for the partitioning procedure, patient-level separation could not be explicitly enforced or retrospectively verified.  The secure variants are compared to study the trade-off between privacy, robustness, verifiability, communication cost, and runtime overhead.

\begin{table}[!t]
\centering
\caption{Experimental setup and protocol parameters.}
\label{tab:experimental_setup}
\small
\begin{tabular}{p{0.17\linewidth}p{0.8\linewidth}}
\hline
\textbf{Category} & \textbf{Setting} \\
\hline
Dataset &
TUH Abnormal EEG Corpus (TUAB), version 3.0.1. The TUAB training split used as the source pool contained 2,717 EDF recordings: 1,371 normal and 1,346 abnormal. These recordings were used to construct configuration-specific client-folder pools for the 10, 40, and 70-client experiments. Data are used only as dataset-level normal/abnormal classes; no clinical claims are made about individual patients. \\

Preprocessing &
EDF files loaded with MNE and resampled to 100Hz; the first 19 signal channels and first 10s (1,000 samples/channel) were retained, with zero-padding when necessary and per-channel standardization. Channels were not explicitly reordered by electrode name. \\

Model and training &
Custom lightweight 1D CNN with three convolutional layers and 60,034
trainable parameters. Local training used cross-entropy loss and the Adam
optimizer with learning rate $5\times10^{-4}$, batch size 16, and two local epochs for 100 FL rounds. The global update parameter in Eq.~(8) was $\eta=1$, so the decoded weighted aggregate was applied directly to the global model.
\\

Client configurations &
For the 10-client configuration, 10 client folders were created with 250 EDF recordings per folder. In each of five runs, 8 clients were used for training and 2 for distributed evaluation, corresponding to 2,000 training and 500 evaluation recordings per run. The training/evaluation client assignment was rotated between runs. For the 40-client configuration, a pool of 80 client folders was created with 30 recordings per folder. In each of five runs, 40 folders were selected, of which 32 were used for training and 8 for distributed evaluation, corresponding to 960 training and 240 evaluation recordings per run. The selected client window was shifted across the 80-folder pool between runs. For the 70-client configuration, a pool of 140 client folders was created with 18 recordings per folder. One 70-client run was performed due to its runtime cost; 56 selected clients were used for training and 14 for distributed evaluation, corresponding to 1,008 training and 252 evaluation recordings. The remaining prepared client folders were not used in the reported run. Fixed client assignments and identical random seeds were not enforced across all protocol variants. 
\\

Protocol settings &
Baseline FL, basic SecAgg comparison, proposed semi-honest setting, proposed malicious setting, semi-honest setting with AN, and malicious setting with AN. The baseline serves as a runtime reference without secure aggregation. \\

Graph and threshold parameters &
For the proposed graph-based variants, graph degree \(k\) was set to 5, 9, and 11 for the 10-, 40-, and 70-client settings, with reconstruction thresholds \(t=3\), \(t=5\), and \(t=6\), respectively. The basic SecAgg comparison used a strict-majority reconstruction threshold \(\lfloor n/2\rfloor+1\), giving thresholds 6, 21, and 36. \\

Common protocol parameters &
Dropout tolerance \(\delta=0.1\), clipping bound \(\tau=0.75\), AN verification vectors \(v=5\), verification modulus \(q=2^{61}-1\), quantization range \(R_Q=2^{22}\), secure-aggregation modulus \(M=2^{32}\), maximum aggregation weight 256 for 10 clients and 64 for 40/70 clients, and model-update dimension 60,034. \\

Experimental scope and limitations &
The 10- and 40-client configurations were evaluated over five runs; the 70-client configuration was evaluated over one run due to runtime cost. No intentional client dropouts, active server equivocation, forged active-set views, malicious-client behavior, or record-linkage evaluation were injected. The Bloom filter-based record-linkage initialization step was bypassed because the TUH-derived data were already anonymized and contained no usable linkage attributes. \\
\hline
\end{tabular}
\vspace{-10pt}
\end{table}

\paragraph{Evaluation metrics.}
Table \ref{tab:mean-results} reports learning behavior and system-level overhead over 100 FL rounds. For the 10-client and 40-client settings, values are averaged over five runs; for the 70-client setting, values are from one run. Agg. reports the mean logged cumulative aggregation-stage time across the 100 FL rounds. Server reports the mean logged cumulative server-side secure-aggregation protocol computation time, including setup, masked-vector handling, dropout recovery, unmasking, and verification computation when applicable. Client reports the mean logged cumulative value of the per-round mean client-side protocol computation time, including key generation, secret sharing, mask construction, masked update preparation, and verification operations. Train and Eval. report mean logged cumulative client-side training and evaluation times. Round reports the mean logged cumulative wall-clock duration across the 100 FL rounds, where each underlying duration was measured from the start of client fitting to the end of round evaluation. Comm. reports the mean logged cumulative estimated secure-aggregation protocol traffic over 100 rounds. It counts serialized protocol payloads exchanged during setup, key/share distribution, masked-update upload, recovery-share exchange, and verification; for AN variants, it also includes AN tag submission and AN public-bundle communication. It does not include full network-layer traffic, raw EEG transfer, or ordinary training/evaluation data loading. Accuracy reports the mean distributed evaluation accuracy over the 100 rounds, averaged over the available runs for each setting.

\begin{table}[!t]
\centering
\caption{Mean cumulative logged runtime, communication, and accuracy values across 100 FL rounds. Accuracy denotes the mean distributed evaluation accuracy across the 100 rounds rather than final-round classification accuracy. The 10-client and 40-client values are averaged over five runs; the 70-client values are from one run.}
\label{tab:mean-results}
\small
\setlength{\tabcolsep}{2.2pt}
\renewcommand{\arraystretch}{1.1}
\begin{tabular}{llrrrrrrrr}
\hline
\textbf{Clients} & \textbf{Protocol} & \textbf{Agg.} & \textbf{Server} & \textbf{Client} & \textbf{Train} & \textbf{Eval.} & \textbf{Round} & \textbf{Comm.} & \textbf{Mean Acc.} \\
 & & \textbf{(s)} & \textbf{(s)} & \textbf{(s)} & \textbf{(s)} & \textbf{(s)} & \textbf{(s)} & \textbf{(bytes)} & \textbf{(\%)} \\
\hline
10 & Base & 0.1912 & N/A & N/A & 401.7239 & 148.9050 & 6309.0174 & --- & --- \\
10 & Malicious & 26.2148 & 2.3421 & 1.6207 & 539.7311 & 170.0401 & 9868.3627 & 256983300 & 79.22 \\
10 & Malicious AN & 50.2738 & 3.0161 & 3.9461 & 538.3105 & 169.8128 & 9893.2166 & 500815700 & 79.82 \\
10 & SecAgg & 11.2530 & 5.3525 & 1.8488 & 539.7365 & 170.0232 & 9855.7303 & 252131900 & --- \\
10 & Semi-honest & 8.7307 & 2.5202 & 0.9297 & 536.9206 & 168.9192 & 9839.9229 & 247596400 & 79.35 \\
10 & Semi-honest AN & 31.9763 & 3.2718 & 3.3066 & 542.2013 & 169.6886 & 9975.2173 & 491431900 & 79.71 \\
\hline
40 & Base & 0.5327 & N/A & N/A & 233.0389 & 109.8832 & 10269.0648 & --- & --- \\
40 & Malicious & 48.2159 & 11.5131 & 2.8732 & 339.2660 & 151.8076 & 16102.4468 & 1096136000 & 68.78 \\
40 & Malicious AN & 121.2449 & 12.1468 & 5.2561 & 347.3689 & 155.0464 & 16569.5135 & 2073281000 & 67.42 \\
40 & SecAgg & 178.6736 & 170.5877 & 10.1733 & 338.4039 & 150.4464 & 16506.7211 & 1117338000 & --- \\
40 & Semi-honest & 25.3120 & 17.5150 & 1.5762 & 338.1106 & 152.1115 & 15983.8985 & 1004901000 & 70.40 \\
40 & Semi-honest AN & 97.7661 & 18.3001 & 3.8937 & 337.7719 & 150.8659 & 16120.2126 & 1982101000 & 77.65 \\
\hline
70 & Base & 0.8845 & N/A & N/A & 233.5986 & 110.8112 & 18014.4685 & --- & --- \\
70 & Malicious & 75.6922 & 23.2744 & 3.5583 & 336.7278 & 150.1089 & 28154.4477 & 2003841000 & 60.29 \\
70 & Malicious AN & 201.9919 & 24.1430 & 5.8994 & 337.5544 & 149.3943 & 28496.2089 & 3717515000 & 66.05 \\
70 & SecAgg & 813.6944 & 802.8229 & 25.2268 & 335.0209 & 145.6566 & 30295.0714 & 2145692000 & --- \\
70 & Semi-honest & 49.4815 & 39.7107 & 1.9470 & 336.7159 & 151.6611 & 27920.0996 & 1771270000 & 58.64 \\
70 & Semi-honest AN & 173.9643 & 40.6820 & 4.3289 & 337.7394 & 149.9747 & 28409.8546 & 3484716000 & 58.66 \\
\hline
\end{tabular}
\par\smallskip
{\footnotesize\noindent --- denotes a value not retained in the experiment logs; ``N/A'' denotes a metric not applicable to baseline federated learning.}
\end{table}

For the 10- and 40-client configurations, variability across the five runs was also examined using the standard deviation of the run-level mean distributed evaluation accuracy. For $\Pi_1$ to $\Pi_4$, respectively, the standard deviations were approximately 0.42\%, 0.38\%, 0.50\%, and 0.53\% with 10 clients, and 1.13\%, 1.20\%, 2.41\%, and 1.34\% with 40 clients. The 70-client configuration was evaluated in a single run, so cross-run variability is not reported for that setting.

\newpage
\paragraph{Results and discussion.}
All runtime and communication values in Table \ref{tab:mean-results} are mean cumulative logged values over 100 FL rounds; the 10-client and 40-client values are averaged over five runs, while the 70-client values are from one run. The communication column reports estimated secure-aggregation protocol traffic, including AN-related messages where applicable, not total network traffic. The baseline has the lowest aggregation time because it does not perform secure aggregation, while the basic SecAgg comparison shows substantially higher overhead as the number of clients increases. Among the proposed secure variants, the semi-honest setting has the lowest secure aggregation time and client-side protocol computation cost. The malicious variant increases overhead because it adds commitment checks, signatures, acknowledgments, and consistency verification. The auxiliary-notary variants further increase communication and protocol computation because they introduce verification tags and notary-related messages. Accuracy values marked with an em dash were not retained in the available experiment logs for the baseline and basic SecAgg comparison runs. This is an experimental limitation, since complete baseline and basic SecAgg accuracy values would allow a more direct learning-performance comparison across all settings. The reported accuracy values are primarily used to verify that iterative federated learning remains functional when the evaluated secure aggregation mechanisms are applied. Numerical accuracy differences between individual secure variants should not be attributed directly to their security or verification mechanisms, since identical file-to-client partitions and random seeds were not enforced across all protocol variants. The accuracy results are therefore interpreted as evidence of learning behavior under the evaluated configurations rather than as a controlled comparison of predictive performance between the security mechanisms. The 10-client and 40-client results are reported as averages over five runs, with the recovered run-level variability in mean distributed evaluation accuracy reported using standard deviations. The 70-client results are based on one run due to runtime cost and should therefore be interpreted cautiously. Accordingly, the 10-client and 40-client results indicate the behavior observed across the repeated runs, while the 70-client results represent only the single evaluated run.

The results show the expected trade-off between protection level and overhead. The semi-honest protocol is the most efficient secure variant, while the malicious variant adds commitment checks, signatures, acknowledgments, and consistency checks that increase protocol cost. The AN variants further increase communication and runtime because verification tags and notary-side aggregation are processed in each round. Overall, the proposed secure-aggregation variants remained compatible with federated model training in the simulated EEG-oriented pipeline, while the malicious and verifiable variants required additional computation and communication resources. The reported configurations cover up to 70 simulated clients; evaluation at larger scales remains future work.

\begin{figure}[t]
\centering
\includegraphics[width=\linewidth]{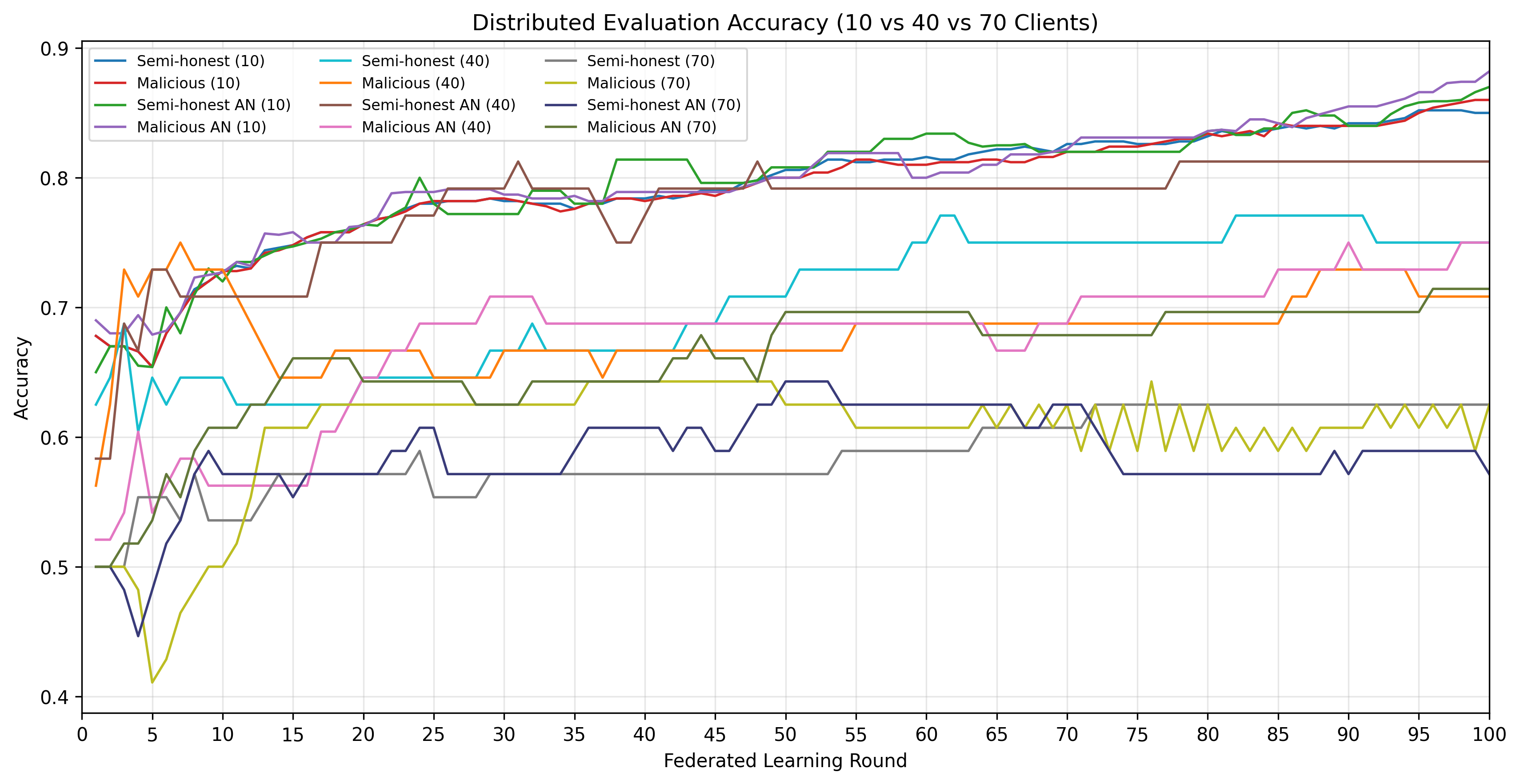}
\caption{Distributed evaluation accuracy across 10-client, 40-client, and 70-client configurations for the secure aggregation variants.}
\label{fig:evaluation-accuracy}
\end{figure}

\begin{figure}[!t]
\centering
\includegraphics[width=\linewidth]{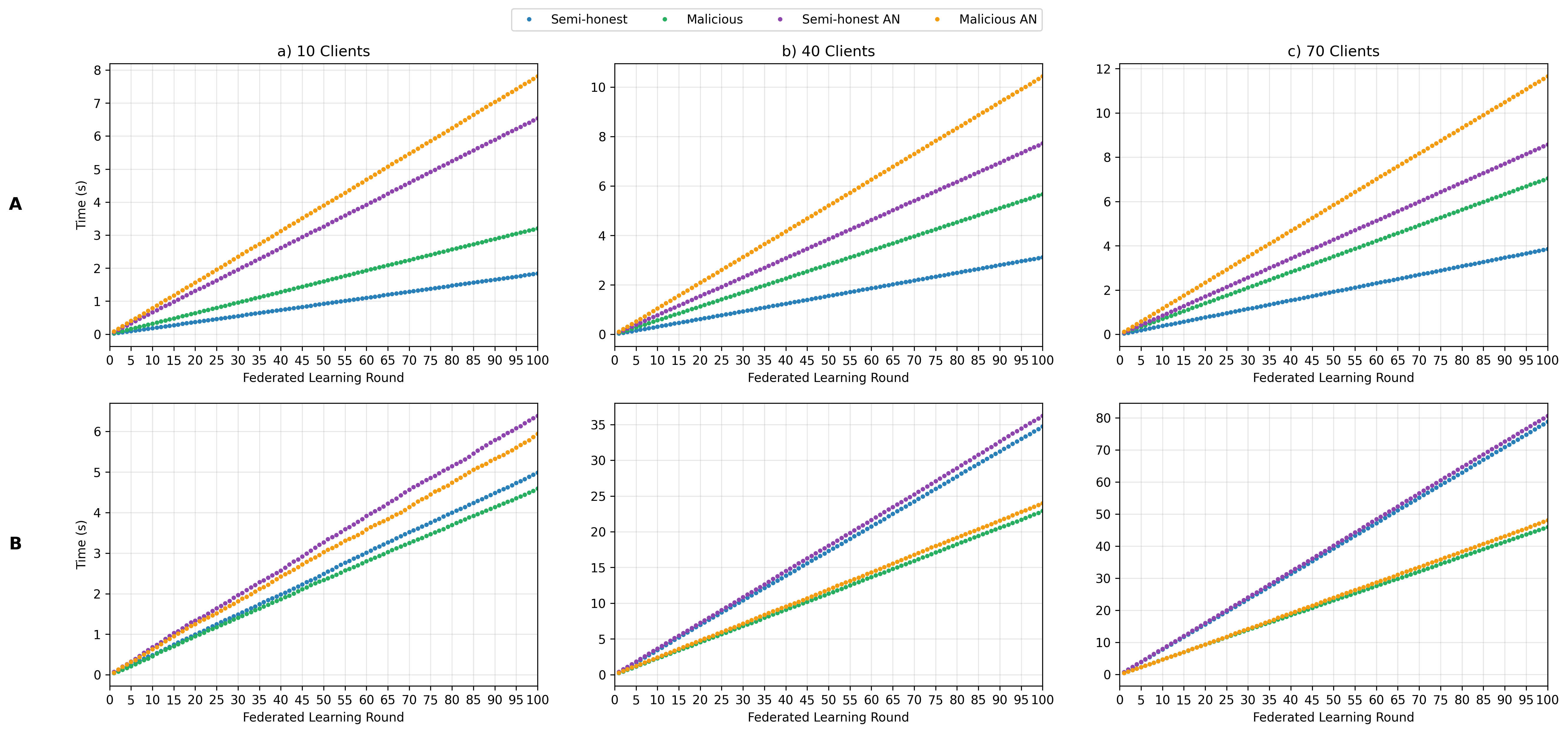}
\caption{Secure-aggregation runtime overhead for the proposed protocols $\Pi_1$--$\Pi_4$ across 10-client, 40-client, and 70-client configurations: (A) client-side overhead and (B) server-side overhead.}
\label{fig:protocol-computation-time}
\end{figure}

Figure~\ref{fig:evaluation-accuracy} shows lower distributed evaluation accuracy in some of the larger-client configurations. This trend should not be attributed to the number of clients alone, since the amount and distribution of training data also differ across configurations. In the 10-client configuration, 8 of the 10 client folders were used for training in each run, with 250 recordings per folder, corresponding to 2,000 training recordings and 500 evaluation recordings. For the 40-client configuration, 40 folders were selected per run from a pool of 80 folders containing 30 recordings each; 32 selected clients were used for training and 8 for distributed evaluation, corresponding to 960 training and 240 evaluation recordings per run. The selected client subset was shifted across the pool between the five runs. For the 70-client configuration, 70 folders were selected from a prepared pool of 140 folders containing 18 recordings each. Of these, 56 clients were used for training and 14 for distributed evaluation, corresponding to 1,008 training and 252 evaluation recordings in the single reported run. Consequently, the observed accuracy differences may reflect the reduced amount of local data per client, differences in the total and distribution of training data, and run-level partition variability, in addition to the number of participating clients. Within each run, the training and evaluation clients were assigned to different client folders, so their EDF recordings did not overlap at the folder level. Across repeated runs, however, client folders could be reused under different training/evaluation assignments as the client selection was rotated. Therefore, the accuracy values should be interpreted as evidence that the privacy-preserving variants remain compatible with useful learning under the available dataset, preprocessing pipeline, model, and FL partitioning, rather than as an optimized EEG classification benchmark. The same fixed 1D-CNN architecture and training configuration are used across all protocol variants as a common reference learning model. Thus, the evaluation compares the secure-aggregation protocols under a consistent learning workload rather than comparing different EEG classification architectures. Differences between AN and non-AN variants are not attributed to the AN mechanism, since AN only verifies aggregate consistency and does not modify accepted updates. Figure \ref{fig:protocol-computation-time} separates client-side and server-side protocol computation. Client-side cost is lowest for the semi-honest variant and increases when malicious-setting checks and AN tag computation are added. Server-side computation generally increases with the number of clients because more messages, shares, recovery operations, unmasking steps, and verification-related values must be processed. Cases where the measured server-side time of the semi-honest variant exceeds that of the malicious variant should be interpreted as implementation-level logging effects, not as evidence that the malicious protocol is cheaper overall, since total overhead is also reflected in client-side computation, communication, and round duration.

\section{Privacy, Security, and Limitations}
The proposed framework is analyzed with respect to the privacy and security objectives introduced by the semi-honest, malicious, and verifiable protocol variants. Table \ref{tab:privacy-security-summary} summarizes the main objectives, the mechanisms used to address them, and the remaining limitations.

\begin{table}[!t]
\centering
\caption{Privacy and security objectives addressed by the proposed framework.}
\label{tab:privacy-security-summary}
\small
\setlength{\tabcolsep}{3pt}
\renewcommand{\arraystretch}{1.08}
\begin{tabular}{p{0.19\textwidth} p{0.38\textwidth} p{0.4\textwidth}}
\hline
\textbf{Objective} & \textbf{Mechanism in the framework} & \textbf{Scope and limitation} \\
\hline
Local data protection &
Raw EEG recordings, clinical labels, preprocessing, and local training remain inside each institution. &
The server receives model updates rather than raw clinical recordings, but FL alone does not remove all privacy risks. \\
\hline
Individual update privacy &
Each client sends a masked update, protected by a self-mask and pairwise masks shared with graph neighbors. &
The server reconstructs only the aggregate over $A'_2$ under the stated threshold and non-collusion assumptions. \\
\hline
Patient identifier protection &
Bloom filter-based record linkage encodes selected identifiers locally using keyed hashing before duplicate resolution. &
Plaintext identifiers are not exchanged, but Bloom filter privacy depends on $len$, $q_g$, $d$, $\gamma$, and key management. \\
\hline
Cross-institution privacy &
Hospitals do not exchange raw EEG data, plaintext identifiers, or individual unmasked updates with each other. &
The final aggregate and global model may still reflect site-level patterns, especially with few clients or strongly non-IID data. \\
\hline
Controlled leakage from released outputs &
Secure aggregation reveals only the aggregate update needed for global model training. &
The final aggregate and global model may still allow inference unless additional mechanisms such as differential privacy are used. \\
\hline
Aggregation correctness &
Pairwise masks cancel in the aggregate, self-masks are removed through reconstruction, and the final sum is computed over $A'_2$. &
Correctness depends on correct mask reconstruction, active-set handling, and enough clients completing the protocol. \\
\hline
Dropout resilience &
Shamir secret sharing allows recovery of self-mask seeds for surviving clients and masking material for dropped clients. &
The protocol tolerates bounded dropout, but aborts if too many clients drop out or too few shares remain available. \\
\hline
Protocol consistency &
Merkle public-key commitments, proof verification, signed inclusion messages, and acknowledgments reduce inconsistent server views. &
These mechanisms limit server equivocation, but do not provide full Byzantine robustness against malicious clients. \\
\hline
Message authenticity &
Authenticated encryption protects share distribution, while signatures protect inclusion and acknowledgment evidence. &
The protection depends on correct key generation, key handling, and signature verification during the required protocol stages. \\
\hline
Aggregate verifiability &
The AN aggregates linear verification tags for the server-declared contributor set, and clients check the released aggregate against these tags. &
The AN is assumed not to collude with the server. The contributor set is not independently validated by the AN; therefore, consistent omission from both the declared set and aggregate is not detected. \\
\hline
\end{tabular}
\vspace{-12pt}
\end{table}

The objectives in Table \ref{tab:privacy-security-summary} are not provided equally by all protocol variants. The semi-honest variant $\Pi_1$ mainly addresses individual update privacy and bounded dropout recovery under an honest-but-curious server. The malicious variant $\Pi_2$ adds consistency and authenticity mechanisms to reduce server equivocation and unsafe recovery-share release. The verifiable variants $\Pi_3$ and $\Pi_4$ add aggregate-level checking through the AN, which compares the released aggregate against verification tags aggregated for the server-declared contributor set. Therefore, the variants provide different combinations of protection, with additional mechanisms increasing protocol complexity and overhead.
In this paper, the malicious setting is limited to secure-aggregation protocol behavior, especially inconsistent server views, unsafe recovery-share release, and active-set manipulation; malicious-client poisoning, backdoors, and Byzantine-robust learning remain outside the scope.

The framework protects raw EEG data by keeping it local, protects individual updates during secure aggregation, supports duplicate handling without plaintext identifier exchange, and improves protocol consistency in the malicious and verifiable variants. The main limitations are that secure aggregation does not prevent all inferences from the final aggregate or global model, and the current design does not fully address poisoning, backdoor, or Byzantine-client attacks. These risks require complementary defenses such as differential privacy, robust aggregation, anomaly detection, or stronger audit mechanisms. The AN also introduces a separate trust assumption, since it is assumed to be independent from the server and receives verification tags that may reveal limited linear information about client updates. As a stronger future extension, direct server-to-AN contributor-set communication could be replaced by client-relayed, server-signed inclusion receipts, allowing the AN to construct and verify the contributor view without relying directly on the server-provided set.

\section{Related Work}
Federated learning has been widely studied as a privacy-preserving approach for collaborative model training without centralizing sensitive data \cite{RiekeH0MRABGLMO20,ThapaliyaOG0CP24}. This is especially relevant in healthcare, where institutional, ethical, and legal constraints often limit direct data sharing. Prior work has applied FL to medical imaging, electronic health records, physiological signals, and EEG-related tasks. In these settings, FL allows hospitals or clinical sites to keep raw patient data local while still contributing to a shared model. However, healthcare FL also introduces practical and security challenges, including non-IID data, heterogeneous local datasets, limited communication budgets, client dropout, and possible privacy leakage from model updates \cite{LiSZSTS20,DingAHAL23}. EEG-oriented FL studies further show that neurophysiological data introduce additional difficulties because of subject variability, noise, and site-dependent signal distributions. Some works have focused mainly on model performance and heterogeneity in EEG or neuroimaging FL, while leaving cryptographic protection of model updates outside the main design. This motivates secure aggregation mechanisms that can be integrated with FL training so that the server learns only an aggregate update rather than individual client contributions.

Secure aggregation is one of the main cryptographic tools used to protect client updates in federated learning. Bonawitz et al. \cite{BonawitzIKMMPRS17} introduced a foundational masking-based secure aggregation protocol that allows a server to compute the sum of high-dimensional client vectors without observing individual updates. Their construction combines pairwise masks generated through key agreement and pseudorandom expansion with self-masks protected by Shamir's secret sharing. The pairwise masks cancel when both clients remain active, while the secret-sharing mechanism allows the server to remove the required masks when some clients drop out. This design is important because it provides the basic structure used by many later protocols: mask each client update locally, recover only the masking material needed for aggregation, and prevent the server from reconstructing an honest surviving client's plaintext update. However, the original construction requires substantial client-to-client key material and communication, especially as the number of clients grows.

Several later protocols improved the scalability and efficiency of masking-based secure aggregation. Bell et al. \cite{BellBGL020} proposed a graph-based secure aggregation protocol, often referred to as SecAgg+, which replaces the dense pairwise masking structure with sparse graph-based communication. Instead of requiring every client to establish masks with every other client, clients communicate only with a limited number of graph neighbors. This reduces the overhead while still supporting dropout resilience under appropriate graph connectivity assumptions. This line of work is especially relevant to our design because our semi-honest protocol also uses graph-neighbor masking and dropout recovery to reduce the cost compared with dense pairwise secure aggregation. LightSecAgg, proposed by So et al. \cite{SoNYL0AGA22}, follows a different direction by using coded mask construction so that the server can reconstruct the aggregate mask of surviving users more efficiently. Rather than reconstructing many pairwise masks, the protocol uses MDS-coded mask shares and moves much of the cost into an offline encoding phase. FastSecAgg \cite{abs-2009-11248} similarly improves scalability by using FFT-based multi-secret sharing to reduce the computational cost of handling high-dimensional updates and many clients. These protocols show that secure aggregation research has moved from basic pairwise masking toward graph-based, coded, and optimized secret-sharing constructions that reduce communication or reconstruction overhead.

Other secure aggregation systems reduce cost by changing how masks are generated, reused, or recovered. SASH \cite{LiuCYFLL22} uses seed-homomorphic pseudorandom generators so that clients can mask updates with single PRG outputs while the protocol separately computes the aggregate masking seed. This avoids dense client-to-client mask cancellation and can simplify deployment. CodedSecAgg \cite{SchlegelKRA23} combines coding techniques with secret sharing so that aggregation can tolerate stragglers and recover the aggregate from a subset of active clients. Flamingo \cite{MaWAPR23} addresses the repeated setup cost that appears in multi-round FL by deriving per-round masking seeds from long-term keys, while still allowing masks to be repaired when clients drop out. DealSecAgg \cite{StockHSD24} introduces a dealer-assisted design that shifts mask recovery away from clients and avoids the dense interaction required by classical masking schemes. These works are not identical to the protocol used in this paper, but they define the broader design space: secure aggregation protocols must balance client overhead, server reconstruction cost, dropout resilience, number of communication rounds, and deployment assumptions.

Although secure aggregation protects individual updates from an honest-but-curious server, it does not automatically guarantee that the server behaves consistently. A server may omit updates, return an incorrect aggregate, manipulate the active-client set, or send inconsistent protocol information to different clients. Prior work has therefore studied attacks and limitations of secure aggregation under stronger adversarial settings. Pasquini et al. \cite{PasquiniFA22} showed that secure aggregation can be weakened when the server manipulates model consistency or protocol views, motivating stronger mechanisms for consistency and validation. Bell et al. \cite{0001GLLM0Y23} also studied secure aggregation together with input validation. Their framework expresses secure aggregation through an encode-mask-decode abstraction and adds ACORN, a zero-knowledge validation layer that allows clients to prove that committed updates satisfy predicates such as range, sparsity, and norm constraints without revealing the updates. This direction is related to our malicious-setting extension because both address limitations of plain secure aggregation when the server or protocol execution cannot simply be assumed to be benign.

Verifiable secure aggregation has been proposed to give clients evidence that the released aggregate is consistent with the submitted information. Li et al. \cite{LiZD25} introduced LVSA, a lightweight verifiable secure aggregation protocol based on linear verification. In LVSA, clients compute inner-product verification values using public randomness, while an auxiliary node aggregates the verification information. The resulting check allows clients to compare the server's released aggregate with independently aggregated verification tags. This approach is important for our work because it provides the main inspiration for the auxiliary-notary mechanism used in our verifiable variants. Our AN component follows the same general idea of using lightweight linear tags, but adapts it to the quantized secure-aggregation domain used in our masking protocol and applies it as an aggregate-consistency check for the active set.

Other verifiable aggregation schemes use different verification mechanisms and trust assumptions. Zhou et al. \cite{zhou2025group} proposed a group-based verifiable secure aggregation protocol where client updates are shared across multiple aggregation groups and verified through linear combination tags. Xu et al. \cite{XuWT25} introduced a compute-node-assisted verifiable aggregation protocol that uses threshold secret sharing, masking, and symmetric MAC-based checks to reduce client overhead under dropout. Behnia et al. \cite{BehniaRECPH24} proposed e-SeaFL, which combines a single-mask aggregation design with authenticated Pedersen vector commitments so that clients can detect server attempts to inject, drop, or modify updates. These works demonstrate that verifiability can be achieved through auxiliary nodes, compute nodes, commitments, MACs, or zero-knowledge techniques. Compared with these approaches, our work uses a lighter auxiliary-notary check that is not intended to provide full adaptive malicious-output protection, but is practical for checking whether the released aggregate is consistent with the verification tags submitted for the active clients.

Alternative privacy-preserving techniques include homomorphic encryption, differential privacy, and multi-party computation \cite{NaeemToorani2026}. Homomorphic encryption-based solutions allow a server to aggregate encrypted model updates and have been studied in healthcare-oriented federated learning settings \cite{TruhnASMKQWGHJLSBBYCHFHKSIBKKNK24,StripelisGSDGASSRNTA24}. Differential privacy instead limits information leakage by perturbing updates or released aggregates, and has been studied for cross-silo federated learning \cite{KatoXTCY24}. Secure aggregation based on secure multi-party computation or replicated secret sharing has also been applied to medical federated learning and robust aggregation \cite{mitrovska2024secure,TangLDMLLDJ25,HuangYZ25}, while secure edge-aggregated healthcare federated learning provides another applied direction \cite{MauryaHPCSSS25}. These methods can provide strong privacy or robustness properties, but often introduce additional computation, communication, infrastructure, or trust assumptions.

\begin{table}[!t]
\centering
\caption{Comparison of representative secure aggregation and verifiable aggregation approaches.}
\label{tab:related-work-comparison}
\small
\setlength{\tabcolsep}{3pt}
\renewcommand{\arraystretch}{1.15}
\begin{tabularx}{\textwidth}{p{0.18\textwidth} p{0.4\textwidth} X}
\toprule
\textbf{Literature} &
\textbf{Main technique} &
\textbf{Malicious / verification support} \\
\midrule

Bonawitz et al. \cite{BonawitzIKMMPRS17} &
Pairwise masking, self-mask, Diffie--Hellman key agreement, and Shamir's secret sharing &
Semi-honest setting; no explicit aggregate verification \\

Bell et al. \cite{BellBGL020} &
Sparse graph-based secure aggregation with reduced client communication &
Semi-honest setting; no explicit aggregate verification \\

So et al. \cite{SoNYL0AGA22} &
Coded mask construction with efficient aggregate-mask reconstruction &
Semi-honest setting; no explicit aggregate verification \\

Kadhe et al. \cite{abs-2009-11248} &
FFT-based multi-secret sharing for efficient secure aggregation &
Semi-honest setting; no explicit aggregate verification \\

Liu et al. \cite{LiuCYFLL22} &
Seed-homomorphic PRG masking with separate masking-key aggregation &
Semi-honest setting; no explicit aggregate verification \\

Schlegel et al. \cite{SchlegelKRA23} &
Coded aggregation combined with one-time padding and secret sharing &
Privacy against colluding parties; no lightweight auxiliary verification \\

Ma et al. \cite{MaWAPR23} &
Reusable per-round masks and threshold decryption for multi-round FL &
Supports recovery from dropped or deviating participants through decryptor-assisted mask recovery \\

Li et al. \cite{LiZD25} &
Linear verification with an auxiliary node and masked aggregation &
Aggregate consistency verification under a non-collusion assumption \\

Behnia et al. \cite{BehniaRECPH24} &
Single-mask aggregation with authenticated Pedersen vector commitments &
Verifies aggregate correctness against server injection, dropping, or modification \\

Bell et al. \cite{0001GLLM0Y23} &
Secure aggregation with zero-knowledge input validation &
Validates committed client updates against predicates such as range, sparsity, and norm constraints \\

Xu et al. \cite{XuWT25} &
Compute-node-assisted masking, threshold sharing, and MAC-based verification &
Verifiable aggregation with compute-node assistance \\

Proposed scheme &
Graph-based masking, quantized integer aggregation, malicious-setting consistency mechanisms, and auxiliary-notary checking &
Supports semi-honest and malicious-setting variants, with optional auxiliary-notary aggregate-consistency verification \\

\bottomrule
\end{tabularx}
\end{table}

Implementation frameworks also affect how secure aggregation can be evaluated in practice. Flower provides a flexible Python-based FL abstraction with customizable client and server behavior, making it suitable for implementing non-standard aggregation workflows. Previous work such as Salvia has implemented SecAgg and SecAgg+ style protocols in Python-oriented FL settings \cite{LiGBL21}. This is relevant because many cryptographic secure aggregation papers focus on protocol design or benchmark simulation, while fewer works integrate several secure aggregation variants into a practical FL framework and evaluate the resulting protocol overhead across different client configurations. To clarify the position of the proposed scheme within this design space, Table \ref{tab:related-work-comparison} compares representative secure aggregation and verifiable aggregation approaches in terms of their main techniques and malicious or verification mechanisms.

\section{Conclusion}
A secure aggregation framework was proposed and evaluated for privacy-preserving federated learning on clinical EEG data. The framework integrated masking-based secure aggregation into a Flower-based cross-silo FL pipeline and combined graph-based communication, threshold secret sharing, dropout recovery, malicious-setting safeguards, auxiliary-notary-based lightweight consistency checking, and an optional Bloom filter-based record-linkage initialization module. The evaluation on TUH EEG-derived data showed that secure variants remained compatible with iterative federated learning, while malicious and verifiable settings introduced additional runtime and communication overhead. Future work will address malicious-client defenses, robust aggregation against poisoning and backdoors, differential privacy, empirical evaluation of the optional record-linkage module when suitable linkage attributes are available, evaluation beyond 70 simulated clients, and optimized cross-silo deployments.

\bibliographystyle{IEEEtran}
\bibliography{Ref}

\end{document}